\documentclass[reprint,aps]{revtex4-1}
\usepackage{amssymb,amscd,amsmath,amsthm}
\usepackage{latexsym,amstext}
\usepackage{dcolumn}
\usepackage{color}
\usepackage{subfigure}
\usepackage{graphicx}
\usepackage{epsfig,epstopdf}

\def\be{\begin{equation}}
\def\ee{\end{equation}}
\def\bea{\begin{eqnarray}}
\def\eea{\end{eqnarray}}
\def\bean{\begin{eqnarray*}}
\def\eean{\end{eqnarray*}}
\def\ds{\displaystyle}
\def\be{\begin{equation}}
\def\ee{\end{equation}}

\def\H{{\mathcal H}}

\def\N{{\mathbb N}}
\def\R{{\mathbb R}}

\def\I{{\mathbb I}}

\def\H{{\mathcal H}}
\def\U{{\mathcal U}}
\def\zf{{\mathcal Z}}
\def\a{{\alpha}}
\def\b{\beta}

\def\l{\langle}
\def\r{\rangle}

\begin{document}

\title{Zernike functions,  rigged Hilbert spaces and potential applications}

\author{E. Celeghini$\,^{\&,}$}
 \email{celeghini@fi.infn.it}
 \affiliation{Dipartimento di Fisica, Universit\`a di Firenze and\\  INFN-Sezione di Firenze
150019
Sesto Fiorentino, Firenze, Italy.}

\author{M. Gadella}
 \email{manuelgadella1@gmail.com}

 \author{M. A. del Olmo}
 \email{marianoantonio.olmo@uva.es}

\affiliation{$^\&$Departamento de F\'{\i}sica Te\'orica, At\'omica y \'Optica and IMUVA \\
Universidad de Valladolid, Paseo Bel\'en 7, 47011 Valladolid, Spain.
}

\date{\today}

\begin{abstract}
We revise the symmetries of the Zernike polynomials that determine the Lie algebra $su(1,1)\oplus su(1,1)$. We show how they induce discrete as well continuous bases that coexist in the framework of rigged Hilbert spaces.  We also discuss some other interesting properties of Zernike polynomials and Zernike functions. One of the interests of Zernike functions has been their applications in optics. Here, we suggest that operators on the spaces of Zernike functions may play a role in optical image processing. 
\begin{description}
\item[PACS numbers]
02.20.Sv, 02.30.Gp, 02.30.Nw, 03.65.Ca, 42.30.Va, 42.30.-d
\end{description}
\end{abstract}


\maketitle
\section{Introduction}\label{introduccion}

The today called Zernike polynomials were introduced by F. Zernike in 1934 \cite{ZER} due to their possible applications in optics.  Nowadays, they are the main ingredient in the construction of the Zernike functions, which are an orthonormal basis for the Hilbert space of square integrable functions on the unit disk. There is a wide bibliography on these functions and their mathematical properties \cite{BWB,kintner1976,lakshminarayanana2011,weisstein,dunkl2001,SP,TO,WUN,ADG,PBKM,GSTL}. Very recent studies underline their role in the analysis of integrable and super-integrable systems as well as the determination of separable coordinates \cite{PSWY,PWY,APWY,PWY2}. It is interesting to remark that Zernike polynomials are the analogues of the spherical harmonics for the disk.

In the optical image processing, adaptive optics serves to clean signals \cite{tyson2015},  where an auxiliary photoreceptor measures the wavefront deformations introduced by the medium and acts on the instrument to induce the opposite effect.  Adaptive optics removes indeed the spurious phases in the complex function $f(r,\phi)$ that represents the optical signal on the disk allowing to obtain a cleaned image $|f(r,\phi)|^2$. This function may not be considered as the final result of the process, but an intermediate step, which may be further elaborated by means of an action that we call {\it soft adaptive optics}.  While ``hard adaptive optics'' acts on the perturbations of the phase introduced solely by the medium, the soft elaboration of the numerical image $|f(r,\phi)|^2$ operates on a set of pixels, so as to obtain another set of pixels. This is independent from the cause of the distortion, i.e., wind in atmosphere, diffraction in lents, defects of the apparatus, etc., and the particular instrument of measure. Hard adaptive optics can only be applied to the cleaning of images, although this soft approach converts images into images and can be employed everywhere. Such a transformation of images can play a role in, for instance, laser physics, microscopic images, radioastronomy or in general instrumental improvement. 

One of the objectives of the present paper is the construction of a theory of operators acting in the space of images defined on the disk. These operators  transform images into images. 
To this end, it is desirable to have an algebra of operators with properties of continuity over some appropriate space. These operators are unbounded with a dense common domain in the space of square integrable functions on the unit disk, $L^2(\mathcal D,r\,dr\,d\phi)$ ($\equiv L^2(\mathcal D)$), where the Zernike functions form an orthonormal basis. Ladder operators should be included  in the algebra of unbounded operators. With this purpose, we need to endow the dense subspace supporting the algebra of operators with a topology stronger than the topology inherited from the Hilbert space $L^2(\mathcal D)$. This leads to the concept of rigged Hilbert space (RHS).

A rigged Hilbert space, also called Gelfand triplet, is a tern of spaces $\Phi\subset\mathcal H\subset \Phi^\times$, where 
$\mathcal H$ is a Hilbert space, $\Phi$ a dense subspace of $\mathcal H$ endowed with a topology finer (with more open sets) than the topology that $\Phi$ has inherit from $\mathcal H$ and $\Phi^\times$ the dual space of $\Phi$. We do not want to discuss properties and other applications of RHS here, since there is a vast available bibliography on the subject \cite{B,R,ANT,M,GG,GG1,GG2}. The space $\Phi$ is the common domain of the operators in the algebra, which with the topology on $\Phi$ become continuous and can be continuously extended into $\Phi^\times$.

In addition, RHS is the proper framework where discrete (complete orthonormal sets in separable Hilbert spaces) and continuous (widely used in quantum mechanics) bases coexist. This is one of the great advantages of RHS, which will  also play a role in our discussion. 

In previous works \cite{CO,CO1,CGO,Cele,Celeghini,CGO1,CGO2}, we have shown the closed relation existing between bases of special functions, Lie algebra representations and RHS. This is also the purpose of the present article, where we shall establish the relations existing between Zernike functions, unitary irreducible representations of $su(1,1)\oplus su(1,1)$, the universal enveloping algebra $UEA[su(1,1)\oplus su(1,1)]$ and our particular choice of RHS. 

This paper is organized as follows:  in Section \ref{zernikesection} we discuss some relevant properties of Zernike functions, while we leave for Section \ref{rigged} a discussion of the RHS implementation for our purposes. In Section \ref{algebra}, we introduce the algebras of continuous operators that will be used in Section \ref{softoptics} to implement the procedure for soft adaptive optics. Finally in the appendices A and B we present some interesting properties of the Zernike polynomials that we have used along this paper. In the appendix C we show a new topology for the space of Zernike functions. This topology is obtained from a family of norms different from the norms used in sections \ref{rigged} and \ref{algebra} but we obtain  similar results those obtained with the original topology.

\section{Zernike functions: discrete and continuous bases on the unit disk}\label{zernikesection}

Zernike functions $\zf_{n}^{m}(r,\phi)$ on the closed unit circle,
\be\label{disk}
\mathcal D=\{(r,\phi)\,, \;\; 0\leq  r\leq 1\,,\; \phi\in [0,2\pi)
\}\,,
\ee
are expressions    of the form  
\be\label{zernike}
\zf_{n}^{m}(r,\phi):=R^m_n(r)\,e^{im\phi}\,,\quad n\in \N\,,\;m\in {\mathbb Z}\,,
\ee
such that
  \be\label{conditions}
 n\in \N\,,\quad m\in {\mathbb Z}\,,\quad
 |m|\le n\,,\quad \frac{n-|m|}{2}\in \N  \,,
 \ee 
  where   $R^m_n(r)$ are real polynomials, called Zernike radial polynomials,   solutions of the following differential equation: 
\begin{widetext}
\begin{equation}\label{1}
\left[ (1-r^2)\,\frac{d^2}{dr^2} -\left(3r-\frac1r   \right) \frac{d}{dr}+n(n+2)-\frac{m^2}{r^2}  \right] R^m_n(r)=0\,,
\end{equation}
\end{widetext}
with  $R^m_n(1)=1$.
 Note that in the differential equation \eqref{1} the label $m$ appear as $m^2$. This shows  that the Zernike polynomials have the symmetry $R^m_n(r)=R^{-m}_n(r)$. Also, it can be seen in the last of \eqref{conditions} that $n$ and $m$ have the same parity (i.e. $n\equiv m$ (mod 2)), in other words $n$ and $m$ ought to be even or odd at the same time.
  An explicit formula for the Zernike polynomials is
  \be\begin{array}{l}\label{Zformula}
\ds R_n^m(r)=\sum_{k=0}^{\frac{n-m}{2}} (-1)^k\,\left( \begin{array}{c} n-k\\ k\end{array}\right)\,\\[0.4cm]
\hskip2cm  \ds \times
\left( \begin{array}{c} n-2k\\ \frac{n-m}{2}- k\end{array}\right)\,r^{n-2k}\,.
\end{array}\ee
Moreover, they are related with the Jacobi polynomials $J_n^{(\a,\b)}(x)$ as follows \cite{BWB,weisstein}
\[
R^m_n(r)= (-1)^{(n-m)/2}\, r^m\, J_n^{(m,0)}(1-2r^2)\,.
\]
Zernike polynomials satisfy some important properties. First of all, for each fixed value of $m$, the polynomials $R^m_n(r)$ fulfill the following orthogonality condition:
\begin{equation}\label{2}
\int_0^1 R^m_n(r)\,R^m_{n'}(r)\,r\, dr= \frac{\delta_{n,n'}}{2(n+1)}\,,
\end{equation}
along a completeness relation of the type
\begin{equation}\label{3}
\sum_{\substack{n=|m| \\ n\equiv m\, {\rm (mod\, 2)}}
} ^\infty  R^m_n(r)\,R^m_n(r')\,(n+1)=\frac{\delta(r-r')}{2r}\,.
\end{equation}

The Zernike polynomials may be extended to any $r\in [-1,1]$. In Appendix 1,  we discuss the main facts relative to this extension and the enlarged symmetries. This generalization gives a relation between   Zernike and  Legendre polynomials.

\subsection{W-Zernike functions}
We redefine the Zernike functions  \eqref{zernike} in a slightly different way  by introducing a numerical factor and by changing the parametrization. We   call  ``W-Zernike functions'' and denote $W_{u,v}(r,\phi)$ to this new  class of functions. 

Let us introduce the parameters $u$ and $v$, defined as \cite{dunkl2001}:
\begin{equation}\label{4}
u:=\frac{n+m}{2}\,,\qquad v:=\frac{n-m}{2}\,.
\end{equation}
The parameters $u$ and $v$ are positive integer numbers  
and are independent of each other.
Hence the $W$-Zernike functions, $W_{u,v}(r,\phi)$ (with $u,v=0,1,2,\dots$) , are  functions on the closed unit circle 
$\mathcal D$ \eqref{disk} defined by:
\begin{equation}\begin{array}{l}
\label{5}
W_{u,v}(r,\phi):=
\ds \sqrt{\frac{u+v+1}{\pi}}\, {\zf}^{u-v}_{u+v}(r,\phi)\\[0.4cm] 
\hskip1cm \ds =\sqrt{\frac{u+v+1}{\pi}}\, R^{|u-v|}_{u+v}(r)\,e^{i(u-v)\phi}\,.
\end{array}
\end{equation}
 Note that when $u$ and $v$ have the same (different)  parity $R^{|u-v|}_{u+v}(r)$ are polynomials of degree even (odd).

The $W$-Zernike functions have the following properties:
\begin{enumerate}

\item
 They  are square integrable in 
 $L^2(\mathcal D,rdr d\phi)\equiv L^2(\mathcal D)$ because the $R^m_n(r)$ are polynomials. 

\item
They satisfy the following identities:
\begin{equation}\label{6}
W_{v,u}(r,\phi)=W_{u,v}(r,\phi)^*= W_{u,v}(r,-\phi)\,,
\end{equation}
where the star denotes complex conjugation. This symmetry property holds from $R^m_n(r)=R^{-m}_n(r)$ and \eqref{5}.

\item
 Orthonormality in $L^2(\mathcal D)$:
\be\begin{array}{l}
\label{7}
\l W_{u',v'},W_{u,v}\r =\\[0.4cm]\ds
 = \int_0^{2\pi} d\phi \int_0^1 dr\,r\, W_{u',v'}(r,\phi)^*\,W_{u,v}(r,\phi)\\[0.4cm]
\ds =\delta_{u,u'}\,\delta_{v,v'}\,.
\end{array}
\ee

\item 
The following completeness relation holds:
\begin{equation}\begin{array}{l}\label{8}
\ds\sum_{u,v=0}^\infty W_{u,v}(r,\phi)\,W^*_{u,v}(r',\phi')\\[0.4cm]
\hskip1cm \ds = \frac{1}{2r}\,\delta(r-r')\,\delta(\phi-\phi')\,.
\end{array}\end{equation}

\item 
 From the property of the Zernike polynomials $\big|R^m_n(r)\big|\leq 1\,,\; 0\leq r\leq 1$\,, 
 we find an upper bound for the Zernike functions
\be\label{upperbound}
\big|W_{u,v}(r,\phi)\big|\le \sqrt{\frac{u+v+1}{\pi}}\,,\quad \forall (r,\phi)\in\mathcal D\,.
\ee
\end{enumerate}


\subsection{Discrete and continuous bases on the unit disk}
The above  properties show that the set of Zernike functions $\{W_{u,v}(r,\phi)\}_{u,v\in \N}$ is an orthonormal basis in $L^2(\mathcal D)$. Hence, for any function $f(r,\theta)\in L^2(\mathcal D)$, we have,   in the sense of convergence on the Hilbert space $L^2(\mathcal D)$, that 
\begin{equation}\label{9}
f(r,\phi)=\sum_{u,v=0}^\infty f_{u,v}\,W_{u,v}(r,\phi)\,,
\end{equation}
where $f_{u,v}$ are complex numbers given by
\begin{equation} \label{10}
f_{u,v}= \int_0^{2\pi} d\phi \int_0^1 dr\,r\, W_{u,v}^*(r,\phi)\,f(r,\phi)\,.
\end{equation}
Moreover, from \eqref{7}, \eqref{8} and \eqref{9} we obtain that
\[\begin{array}{lll}
\l f,f\r&=&
\ds \int_0^{2\pi} d\phi \int_0^1 dr\,r\, f^*(r,\phi)\,f(r,\phi)\,
\\[0.3cm] &=&\ds \sum_{u,v=0}^\infty |f_{u,v}|^2<\infty\,.
\end{array}\]

In adaptive optics, one always chooses  
$f(r,\phi)$ real, so that $f_{u,v}=f^*_{v,u}$. 

As is customary in quantum mechanics, let us introduce the generalized continuous basis $\{|r,\phi\rangle\}_{(r,\phi)\in\mathcal D}$ whose elements have the following properties:
\begin{equation}\begin{array}{l}\label{11}
\ds \langle r,\phi|r',\phi'\rangle=\frac1r\,\delta(r-r')\,\delta(\phi-\phi')\,,\\[0.3cm] 
\ds\mathbb I=\int_0^{2\pi} d\phi \int_0^1 dr\,r\, |r,\phi\rangle\langle r,\phi|\,,
\end{array}\end{equation}
where $\mathbb I$ is the identity operator. Next, we define the kets $|u,v\rangle$ with the help of the  Zernike functions $W_{u,v}(r,\phi)$
\begin{equation}\label{12}
|u,v\rangle:= \int_0^{2\pi} d\phi \int_0^1 r\,dr\, |r,\phi\rangle\,W_{u,v}(r,\phi)\,,
\end{equation}
with $u,v=0,1,2,\dots\,,$, 
which have the following properties as one can check from the ortogonality relations \eqref{7} and \eqref{11}:
\begin{equation}\begin{array}{l}\label{13}
\ds \langle u,v|u',v'\rangle=\delta_{u,u'}\,\delta_{v,v'}\,, \\[0.3cm]
\ds \sum_{u,v=0}^\infty |u,v\rangle\langle u,v|=\mathcal I\,.
\end{array}\end{equation}
Due to the fact that $\{|u,v\rangle\}_{u,v\in \N}$ is a discrete basis and $\{|r,\phi\rangle\}_{(r,\phi)\in\mathcal D}$ a continuous basis, we make a distinction between the identities $\mathbb I$ and $\mathcal I$, which, in principle, should be different. 
From  expression (\ref{12}) and taking into account the first relation of (\ref{11}), we obtain
\begin{equation}\label{14}
\langle r,\phi|u,v\rangle= W_{u,v}(r,\phi)\,,
\end{equation}
so that the set of vectors $\{|u,v\rangle\}$ forms an orthonormal basis on a Hilbert space, $\mathcal H$, unitarily equivalent to $L^2(\mathcal D)$ and the Zernike functions are the transition elements between both bases.  On this Hilbert space $\mathcal H$  the identity operator is $\mathcal I$ \eqref{13}. 

Taking into account relations \eqref{9}, \eqref{10} and \eqref{14} we have  
for any $f(r,\phi)\in L^2(\mathcal D)$ \eqref{9} 
that
\[
|f\r:=\sum_{u,v=0}^\infty f_{u,v}\,|u,v\r\,
\in \mathcal H\,.
\]
Then, the space of all vectors 
\be\label{f}
|f\rangle=\sum_{u,v=0}^\infty f_{u,v}\,|u,v\r
\ee
such that $\ds \sum_{u,v=0}^\infty |f_{u,v}|^2<\infty$
completes an abstract Hilbert space, $\mathcal H$, and the mapping 
\begin{equation}\label{15}\begin{array}{llll}
\mathcal U\,:\,&{\mathcal H}\;\;&\longmapsto \;\; & L^2(\mathcal D)\\[0.3cm]
&|f\r\;\;&\longmapsto \;\; &{\mathcal U}|f\rangle=\langle r,\phi | f\rangle=f(r,\phi)
\end{array}\end{equation}
 is unitary. The vectors $|f\rangle\in\mathcal H$ admit two representations in terms of the continuous basis and the discrete basis, respectively, 
\[ \label{16}
\begin{array}{lll}
|f\rangle &=&\ds \int_0^{2\pi} d\phi \int_0^1 dr\,r\,|r,\phi\rangle\langle r,\phi|f\rangle
 \\[2ex]
 &=& \ds\int_0^{2\pi} d\phi \int_0^1 dr\,r\,|r,\phi\rangle\,f(r,\phi)\,, \\[2ex]
|f\rangle &=&\ds \sum_{u,v=0}^\infty |u,v\rangle\langle u,v|f\rangle=\sum_{u,v=0}^\infty |u,v\rangle\,f_{u,v}\,.
\end{array}\]
Although the first of these two relations gives the unitary mapping $\mathcal U$, as a matter of fact  (\ref{15}) is strictly valid on a dense subspace of $\mathcal H$. The second is just the span of $|f\rangle\in\mathcal H$ with respect to the basis $\{|u,v\rangle\}$. These two spans provide of two expressions for the scalar product of any two vectors $|g\rangle,|f\rangle\in\mathcal H$ as well as the norm of any vector  $|f\rangle\in\mathcal H$ as
\[\begin{array}{lll}
\langle g|f\rangle &=&\ds
\int_0^{2\pi} d\phi \int_0^1 dr\,r\, g(r,\phi)^*\,f(r,\phi)\\[2ex]
&=&\ds\sum_{u,v=0}^\infty g^*_{u,v}\,f_{u,v}\,, \label{17}
\end{array}\]
\[\begin{array}{lll}
||f||^2 :=\langle f|f\rangle &=&\ds \int_0^{2\pi} d\phi \int_0^1 dr\,r \,|f(r,\phi)|^2\\[2ex]
&=&\ds\sum_{u,v=0}^\infty |f_{u,v}|^2\,.\label{18}
\end{array}
\]
Note that, according to (\ref{15}), we have for $|f\rangle\in\mathcal H$   that 
\begin{equation}\label{19}
\langle r,\phi|f\rangle=f(r,\phi)=(\mathcal Uf)(r,\phi)\,.
\end{equation}

As already noted, the first identity in (\ref{19}) is not valid for any $|f\rangle\in\mathcal H$, but only for those on a dense subspace of $\mathcal H$. We shall clarify this point later. 

\section{A proposal for Rigged Hilbert Spaces}\label{rigged}

To begin with, let us consider the set $\Phi_1\subset \H$  of vectors $|f\rangle$ \eqref{f} such that
\begin{equation}\label{21}
\big|\big|f\rangle\big|\big|_p^2:=\sum_{u,v=0}^\infty \big|f_{u,v}\big|^2\,(u+v+1)^{2p}<\infty\,,
\end{equation}
for any $p=0,1,2,\dots\,.$
This is a countable normed subspace, and hence metrizable, of $\mathcal H$. Its norms ($\big|\big|-\big|\big|_p\,,\; p=0,1,2,\dots$) are given by (\ref{21}). Then, consider the subspace $\Psi_1:=\mathcal U\Phi_1$ of $L^2(\mathcal D)$. Hence
$ \Psi_1$ is the set of $ f(r,\phi)\in L^2(\mathcal D)$ \eqref{9}
such that 
\[
\sum_{u,v=0}^\infty \big|f_{u,v}\big|^2\,(u+v+1)^{2p}<\infty\,,\;\;\; \forall p\in N  .
\]
It is obvious that $\Psi_1=\mathcal U\Phi_1$ has the metrizable structure transported from $\Phi_1$ by $\mathcal U$.

Next, we consider a subspace $\Psi$ of $\Psi_1$ with the following additional conditions:
\begin{itemize}
\item[i)]
 The series \eqref{9}, i.e. 
\[
f(r,\phi)=\sum_{u,v=0}^\infty f_{u,v}\,W_{u,v}(r,\phi)\,,
\]
converges pointwise almost elsewhere   in $\mathcal D$. Note that in general, $L^2$ convergence does not imply pointwise convergence. 

\item[ii)]
 If $f(r,\phi)\in\Psi$, then, $r\,e^{i\phi}f(r,\phi)\in\Psi$. 

\end{itemize}
Condition i) is satisfied by all finite linear combinations of Zernike functions $W_{u,v}(r,\phi)$. 

In order to prove  condition ii) let us start by 
defining   the operator  $P$ 
\be\label{p}
Pf(r,\phi):=r\,e^{i\phi}\,f(r,\phi)\,.
\ee
 Then, $P$ transforms the Zernike functions into linear combinations of (only two) Zernike functions as (see  the Property 2 in the Appendix 2 for the proof)
\begin{equation}\begin{array}{l}\label{23}
P\,W_{u,v}(r,\phi)=\a_{u}^{v}\,W_{u+1,v}(r,\phi)+\b_{u}^{v}\,W_{u,v-1}(r,\phi)
\end{array}\end{equation}
where the coefficientes $\a_{u}^{v}$ and $\b_{u}^{v}$ are given by
\be\begin{array}{l}\label{Wcoeff}
\ds\a_{u}^{v}=\frac{u+1}{\sqrt{(u+v+1)(u+v+2)}}\,,\\[0.3cm]
\ds\b_{u}^{v}=\frac{v}{\sqrt{(u+v)(u+v+1)}}
\end{array}\ee
Note that $0\le \a_u^v\le 1$ and $0\le \b_u^v\le 1$.

The consequence is that all finite linear combinations of Zernike functions satisfy condition ii). Since finite linear combinations of elements in an orthonormal basis form a dense subspace of the Hilbert space, we must conclude that $\Psi$ is dense in $L^2(\mathcal D)$. 
Then $\Phi:=\mathcal U^{-1}\Psi$, which is dense in $\mathcal H$. 

We have the following sequence of spaces:
\begin{equation}\label{26}
\Phi\subset\Phi_1\subset \mathcal H\subset \Phi^\times_1\subset\Phi^\times\,,
\end{equation}
where $\Phi^\times$ is the antidual space of $\Phi$. We denote the action of $F\in\Phi^\times$ on any $|f\rangle\in\Phi$ as $\langle F|f\rangle$ and this notation will be kept for the action of $F\in\Phi_1^\times$ on $|f\rangle\in\Phi_1$. Note that $\Phi$ should not necessarily be a closed subspace of $\Phi_1$ (we do not have any proof thereof) and it may well happen that $\Phi^\times_1=\Phi^\times$. In any case, this is not relevant in our discussion. The space $\Phi$ will always be endowed with the topology inherited from that of $\Phi_1$, i.e., the metrizable topology given by the countable set of norms (\ref{21}). 

Along to the spaces \eqref{26}, we have their representations which are their images by $\mathcal U$. Note that if we have $\Phi\subset \mathcal H\subset\Phi^\times$, we may extend $\mathcal U$ to $\Phi^\times$ by using the {\it duality formula}:
\begin{equation}\label{27}
\langle \mathcal U\,F|\,(\mathcal U\,|f\rangle)\rangle:=\langle F|f\rangle \,,
\end{equation} 
valid for any $|f\rangle\in\Phi$ and any $F\in\Phi^\times$. This defines $\mathcal U$ on $\Phi^\times$(we have denoted the extension also by $\mathcal U$) and the same formula is valid to define $\mathcal U$ on $\Phi_1^\times$. Since $\Psi\equiv \mathcal U\,\Phi$ and $\Psi_1\equiv \mathcal U\,\Phi_1$ and the respective topologies are those transported by $\mathcal U$, which is one to one and onto in both cases, it results that $\Psi^\times \equiv \mathcal U\,\Phi^\times$ and $\Psi_1^\times \equiv \mathcal U\,\Phi_1^\times$. Equivalently to the chain of spaces \eqref{26}, we have another sequence
\begin{equation}\label{28}
\Psi\subset \Psi_1\subset L^2(\mathcal D)\subset \Psi_1^\times\subset\Psi^\times\,.
\end{equation}
This chain of spaces given may be looked as a representation of \eqref{26} by means of the mapping $\mathcal U$.  In other words, we have the diagram
\[
{\cal U}\hskip-0.45cm\begin{array}{lllllclllll}
&\Phi&\subset &\Phi_1&\subset &{\mathcal H}&\subset &\Phi^\times_1&\subset &\Phi^\times
\\[0.2cm]
 & \downarrow & &\hskip-0.30cm {\U}\downarrow && \hskip-0.30cm {\U}\downarrow &&\hskip-0.30cm {\U}\downarrow && \hskip-0.30cm {\U}\downarrow 
\\[0.2cm]
&\Psi &\subset &\Psi_1 &\subset & L^2(\mathcal D)&\subset & \Psi_1^\times&\subset &\Psi^\times
\end{array}\,.
\]
While vectors in $\Phi$, $\Phi_1$ and $\H$ are abstract objects, vectors in $\Psi$, $\Psi_1$ and 
$L^2(\mathcal D)$ are square integrable functions on the unit circle.

Before proceeding with our discussion, let us recall an important result concerning continuity of linear mappings on countably normed spaces as those under our consideration. Assume that the topology of an infinite dimensional vector space $\Phi$ is given by the countable set of norms $\{||-||_n\}_{n\in\mathbb N}$ on $\Phi$. Then \cite{RSI}:
\begin{enumerate}
\item 
 A linear functional $F:\Phi\longmapsto\mathbb C$, where $\mathbb C$ is the field of complex numbers, is continuous if and only if, there exists a constant $K>0$ and a finite collection of norms $\left\{\big|\big|-\big|\big|_{n_1},\big|\big|-\big|\big|_{n_2},\dots,\big|\big|-\big|\big|_{n_k}\right\}$ such that for all $f\in\Phi$, we have that
\begin{equation}\label{29}
\big|F(f)\big|\le K\,(\big|\big|f\big|\big|_{n_1}+\big|\big|f\big|\big|_{n_2}+\dots+\big|\big|f\big|\big|_{n_k})\,.
\end{equation}

\item
A linear mapping $A:\Phi\longmapsto \Phi$ is continuous if and only if for any norm $||-||_n$, there exists a positive constant $K_n>0$ and $k(n)$ other norms ($K_n$ and $k(n)$ will depend in general of $n$), such that for any $f\in\Phi$, we have:
\begin{equation}\label{30}
||Af||_n\le K_n (||f||_1+\dots+||f||_{k(n)})\,
\end{equation}
for any $n=1,2,\dots$.
\end{enumerate}

Let us go back to our general discussion. For $r\in[0,1]$ and $\phi\in[0,2\pi)$ fixed, let us define the functional $\langle r,\phi|$ on $\Phi$ as:
\begin{equation}\label{31}
\langle r,\phi|f\rangle:= \mathcal U\,|f\rangle=f(r,\phi)\,.
\end{equation}
Then, we have the following\medskip

\noindent
{\bf Proposition 1.-} {\sl The functional $\langle r,\phi|$ is continuous on $\Phi$ for all $r\in[0,1]$ and $\phi\in[0,2\pi)$.}
\medskip

\noindent 
{\bf Proof.-} The functional is obviously well defined. Taking into account the upper-bound for the Zernike functions \eqref{upperbound} and  by our hypothesis the series (\ref{9}) converges  pointwise almost elsewhere, we have that:
\begin{widetext}
\be\begin{array}{l}\label{32}
|\langle r,\phi|f\rangle|=\ds|f(r,\phi)|\le \sum_{u,v=0}^\infty |f_{u,v}|= \sum_{u,v=0}^\infty \frac{|f_{u,v}|\,(u+v+1)}{(u+v+1)}  \\[0.5cm]
\hskip1cm \ds \le \sqrt{\sum_{u,v=0}^\infty |f_{u,v}|^2\,(u+v+1)^2}\,\sqrt{\sum_{u,v=0}^\infty\frac1{(u+v+1)^2}} = K\,||f||_1\,,
\end{array}\ee
\end{widetext}
with $||\,|f\rangle||_1$ as given in (\ref{21}) for $p=1$ and $K$ is the second square root in the second row of (\ref{32}). This shows the continuity of $\langle r,\phi|$ on $\Phi$.\hfill  $\blacksquare$\medskip

On the other hand since $\Phi$ is metrizable, this implies that $\langle r,\phi|$ could be continuously extended to the closure of $\Phi$, if it were not closed in $\Phi_1$.

Note that, in particular $\langle r,\phi|u,v\rangle= W_{u,v}(r,\phi)$ (\ref{14}) holds and  is well defined now. 

For the  operator $P$ \eqref{p}  
we can prove that for any $ r\in[0,1]$ and any $\phi\in[0,2\pi)$
\begin{equation}\label{34}
P\,|r,\phi\rangle= r\,e^{-i\phi}\,|r,\phi\rangle\, .
\end{equation}
Effectively, using the duality formula (with the same symbol $P$ for either $P$ on $L^2(\mathcal D)$ and $U^{-1}PU$ on $\mathcal H$) the definition \eqref{31} and that $\langle r,\phi|f\rangle= \langle f|r,\phi\rangle^*$ we get that
\begin{equation}\begin{array}{l}\label{33}
\langle f|P|r,\phi\rangle=\langle Pf|r,\phi\rangle =\langle r,\phi|Pf\rangle^*\\[0.3cm]
\hskip0.5cm=[r\,e^{i\phi}\,f(r,\phi)]^*=r\,e^{-i\phi}\langle f|r,\phi\rangle\,.
\end{array}\end{equation}
Omitting the arbitrary $|f\rangle\in\Phi$, we prove the result \eqref{34}.
Note that $P$ is bounded on both Hilbert spaces, but that we do not have any conclusion about the continuity or not of $P$ on $\Phi$. 

\section{Algebras of continuous operators on $\Phi_1$}\label{algebra}

Now, the idea is to show that $\Phi_1$ serves as support of a Lie algebra and its generators are continuous and essentially self-adjoint.   
\subsection{Operators on $\Phi_1$  and $L^2(\mathcal D)$}

To begin with, let us define the operators $U$ and $V$ on $\Phi_1$
\begin{equation}\label{35}
U|u,v\rangle:=u\,|u,v\rangle\,,\quad V|u,v\rangle:=v\,|u,v\rangle\,.
\end{equation}
Then, one may define for $f\in\Phi_1$ (see \eqref{f} and \eqref{21}) 
\begin{equation}\begin{array}{lll}\label{36}
\ds U\,|f\rangle&=&\ds \sum_{u,v=0}^\infty u\,f_{u,v}\,|u,v\rangle\,,\\[0.3cm] 
\ds V\,|f\rangle&=&\ds \sum_{u,v=0}^\infty v\,f_{u,v}\,|u,v\rangle\,.
\end{array}\end{equation}

\noindent
{\bf Proposition 2.-} {\sl The operators  $U$ and $V$ are continuous and essentially self-adjoint on $\Phi_1$.} \medskip

\noindent 
{\bf Proof.-} 
Take for instance,
\[\begin{array}{l}\label{37}
\ds ||U\,|f\rangle||_p^2=\sum_{u,v=0}^\infty |f_{u,v}|^2\,u^2(u+v+1)^{2p}\\[0.3cm]
\hskip0.5cm \ds \le \sum_{u,v=0}^\infty |f_{u,v}|^2\,(u+v+1)^{2p+2} =||\,|f\rangle||_{p+1}^2\,,
\end{array}
\]
which proves the continuity of $U$. In order to see that $U$ is essentially self-adjoint on $\Phi_1$, note that it is symmetric (Hermitian) on $\Phi_1$. Then,  let us define the vectors
\begin{equation}\label{38}
|g_\pm\rangle:= \sum_{u,v=0}^\infty\,\frac{f_{u,v}}{u\pm i}\,|u,v\rangle\,.
\end{equation} 
Observe that
\[\begin{array}{l}\label{39}
\ds ||\,|g_\pm\rangle||_p^2= \sum_{u,v=0}^\infty \,\frac{|f_{u,v}|^2}{u^2+1}\,(u+v+1)^{2p} \\[0.3cm]
\ds \hskip0.5cm\le \sum_{u,v=0}^\infty \, |f_{u,v}|^2\,(u+v+1)^{2p}=||\,|f\rangle||^2_p\,.
\end{array}\]
Hence, $|g_\pm\rangle\in\Phi_1$. Thus, $(U\pm iI)\Phi_1=\Phi_1$, which is dense in $\mathcal H$. Therefore, $U$ is essentially self-adjoint with domain $\Phi_1$. 

The proof for the identical results referred to $V$ is similar.\hfill $\blacksquare$\medskip

Let us define on appropriate dense subspaces of $L^2(\mathcal D)$ the following operators:

\begin{enumerate}

\item 
$d/d{\mathcal R}$: derivation with respect to the radial variable $r$. 

\item $\mathcal R$: multiplication by the radial variable $r$.

\item 
 $e^{\pm i\tilde\Phi}$: multiplication by $e^{\pm i\phi}$, where $\phi$ is the angular variable.
\end{enumerate}

Correspondingly, we have the following operators on dense domains in $\mathcal H$:
\begin{enumerate}

\item[4.]
$D_R:= U\,(d/d{\mathcal R})\,U^{-1}$.

\item[5.]
 $R=U\,\mathcal R\,U^{-1}$.

\item[6.] 
$e^{\pm i\Phi}=U\,e^{\pm i\tilde\Phi} \,U^{-1}$. 
\end{enumerate}

Thus, we have the following formal operators, that are symmetries of the Zernike functions (see \cite{Cele,Celeghini} and Appendix 1) on $\mathcal H$:
\begin{widetext}
\[\begin{array}{l}
 A_\pm=\ds  \frac{e^{\pm i\Phi}}{2} \left[ \mp(1-R^2)\,D_R+R(U+V+1\pm1)+\frac 1R (U-V)  \right]\;\sqrt{\frac{U+V+1\pm 1}{U+V+1}}\,,
\\[0.4cm]
 B_+= \ds \frac{e^{\mp i\Phi}}{2} \left[ \mp(1-R^2)\,D_R+R(U+V+1\pm 1)-\frac 1R (U-V)  \right]\;\sqrt{\frac{U+V+1\pm 1}{U+V+1}}\,.
\end{array}\]
\end{widetext}
The operators $\mathcal A_\pm:=UA_\pm U^{-1}$, $\mathcal B_\pm:=UB\pm U^{-1}$ with dense domain on $L^2(\mathcal D)$, have the following properties:
\be\begin{array}{l}\label{44}
\mathcal A_+\,W_{u,v}(r,\phi)= (u+1)\, W_{u+1,v}(r,\phi)\,,  \\[2ex]
 \mathcal A_-\,W_{u,v}(r,\phi)= u\,W_{u-1,v}(r,\phi)\,,\\[2ex]
\mathcal  B_+\, W_{u,v}(r,\phi)= (v+1)\,W_{u,v+1}(r,\phi)\,,\\[2ex]
 \mathcal B_-\,W_{u,v}(r,\phi)= v\,W_{u,v-1}(r,\phi)\,,  \\[2ex]
\end{array}\ee
Therefore,
\[\begin{array}{l}
A_+\,|u,v\rangle= (u+1)\,|u+1,v\rangle\,,\\[2ex] A_-\, |u,v\rangle=u\,|u-1,v\rangle\,,\label{46}\\[2ex]  
B_+\,|u,v\rangle=(v+1)\,|u,v+1\rangle\,,\\[2ex] B_-|u,v\rangle=v\,|u,v-1\rangle\,.
\end{array}\]

\noindent
{\bf Proposition 3.-} {\sl The operators $A_\pm$ and $B_\pm$ are continuous. Furthermore 
 $A_\pm $ are formal adjoint of each other and the same is true for $B_\pm$\,.}
\medskip

\noindent {\bf Proof.-}  It is quite similar to the proof showing same properties for $U$ and $V$. It is sufficient to give it for one case, say $A_+$. Take $|f\rangle\in\Phi_1$ as in  \eqref{f} and \eqref{21}. Then,
\begin{equation}\label{48}
A_+\,|f\rangle=\sum_{u,v=0}^\infty f_{u,v} (u+1)\,|u+1,v\rangle\,,
\end{equation}
so that
\[\begin{array}{l}\label{49}
||A_+\,|f\rangle||_p^2 =\ds \sum_{u=1,v=0}^\infty |f_{u,v}|^2 \,(u+1)^2\,(u+v+1)^{2p}\nonumber\\[2ex]
\hskip0.5cm\le \ds\sum_{u,v=0}^\infty |f_{u,v}|^2\,(u+v+1)^{2(p+1)}=||\,|f\rangle||^2_{p+1}\,,
\end{array}\]
which proves the continuity on $\Phi_1$. \hfill $\blacksquare$

\subsection{The Lie algebra $su(1,1)\oplus su(1,1)$}

On $\Phi_1$, we have the following commutation relations:
\begin{equation}\label{50}
[U,A_\pm]=\pm A_\pm\,,\qquad [V,B_\pm]=\pm B_\pm\,.
\end{equation}
Then, let us define
\begin{equation}\label{51}
A_3:= U+\frac 12\,,\qquad B_3:= V+\frac 12\,,
\end{equation}
so that we have the following commutation relations
\begin{equation}\begin{array}{ll}\label{52}
[A_+, A_-]=-2 A_3\,,\qquad & [ A_3, A_\pm]=\pm  A_\pm\,,\\[0.3cm]
 [ B_+,  B_-]=-2  B_3\,,\qquad & [  B_3,  B_\pm]=\pm   B_\pm\,,
\end{array}\end{equation}
 showing  that the operators $A_\pm,A_3$  on one side and $B_\pm,B_3$ on the other close  $su(1,1)$ Lie algebras with Casimir invariants, respectively,
 \[\begin{array}{l}\label{54}
{\mathcal C}_{A}=  A_3^2- \frac 12\,\{A_+,A_-\}\;\; \Rightarrow \;\; {\mathcal C}_{A}\,|u,v\rangle=-\frac 14\,|u,v\rangle\,,
\\[0.3cm]
{\mathcal C}_{B}= B_3^2-\frac 12\,\{B_+,B_-\} \;\; \Rightarrow \;\; {\mathcal C}_{B}\,|u,v\rangle=-\frac 14\,|u,v\rangle\,.
 \end{array}\]
where $\{X, Y\}$ denotes the anticommutator of the operators $X$ and $Y$, i.e.,   $\{X, Y\}:= X\, Y+Y\, X$.
We may easily check that all $A$-operators commute with all $B$-operators, i.e., 
\begin{equation}\label{53}
[A_i,B_j]=0\,,\qquad i,j=+,-,3\,.
\end{equation} 
 Thus with the $A$ and $B$-operators   we have obtained a realization of the six dimensional Lie algebra $su(1,1)\oplus su(1,1)$  recovering previous results by \cite{WUN}.
 
We may compare this result with the Casimir for the discrete principal series of unitary irreducible representations for the group $SU(1,1)$, which is given by ${\mathcal C}=j(j-1)\I$, with $j=1/2,\,1,\,3/2,\dots$ \cite{LN}. Here, $j(j-1)=-1/4$ and therefore, $j=1/2$. The space supporting this representation is usually denoted as $D^+_{1/2}$, so that the space spanned by the Zernike functions must be isomorphic to the space $D^+_{1/2}\otimes D^+_{1/2}$, which supports an irreducible unitary representation of the group $SU(1,1)\otimes SU(1,1)$. Its corresponding Lie algebra is spanned by six operators $\{A_\pm,B_\pm,A_3,B_3\}$ that act on a basis, $|a,b\rangle$, of $D^+_{1/2}\otimes D^+_{1/2}$ as:
\be\begin{array}{l}\label{56}
A_\pm \,|a,b\rangle=(a\pm \frac12)\,|a\pm 1,b\rangle\,,\\[2ex] A_3\,|a,b\rangle= a\,|a,b\rangle\,, \\[2ex]  B_\pm \,|a,b\rangle=(b\pm \frac12)\,|a,b\pm 1\rangle\,, \\[2ex] B_3\,|a,b\rangle= b\,|a,b\rangle\,.
\end{array}\ee
There is an immediate relation between $|a,b\rangle$ and $|u,v\rangle$ and is given by $u=a+1/2$ and $v=b+1/2$. 
\subsection{The universal enveloping algebra of $su(1,1)\oplus su(1,1)$}

Now, let us call UEA$[su(1,1)\oplus su(1,1)]$ to the universal enveloping algebra of $su(1,1)\oplus su(1,1)$. This is the vector space spanned by the ordered monomials of the form $A_+^{\alpha_1}\,A_3^{\alpha_2}\,A_-^{\alpha_3}\,B_+^{\beta_1}\,B_3^{\beta_2}\,B_-^{\beta_3}$, where $\alpha_i$ and $\beta_j$, $i,j=1,2,3$ are either zero or natural numbers (see the Poincar\'e-Birkoff-Witt theorem \cite{varadarajan}). If we denote by $\overline \alpha=(\alpha_1,\alpha_2,\alpha_3)$ and $\overline \beta=(\beta_1,\beta_2,\beta_3)$, any operator $ O\in {\rm UEA}[su(1,1)\oplus su(1,1)]$ has the following form: 
\be\begin{array}{l}\label{58}
 O=\ds \sum_{\overline\alpha,\overline\beta} O_{\overline\alpha,\overline\beta} \\[0.3cm]
 \hskip0.5cm  
 \ds =\sum_{\overline\alpha,\overline\beta} c_{\overline\alpha,\overline\beta}\, A_+^{\alpha_1}\,A_3^{\alpha_2}\,A_-^{\alpha_3}\,B_+^{\beta_1}\,B_3^{\beta_2}\,B_-^{\beta_3}\,,
\end{array}\end{equation}
where $c_{\overline\alpha,\overline\beta}$ are complex numbers. 

The unitary mapping $\mathcal U:\mathcal H\longmapsto L^2(\mathcal D)$ transforms this abstract representation into a differential representation  of  the algebra $su(1,1)\oplus su(1,1)$  supported on $L^2(\mathcal D)$. In fact, Zernike functions satisfy the same relation of the Zernike radial polynomials $R_n^m(r)$. Indeed \eqref{1} can be rewritten as
\begin{widetext}
\[
\label{59}
\ds\frac{d^2}{dr^2}\,W_{u,v}(r,\phi)
= \frac1{1-r^2}\;\left[ \left( 3r-\frac1r  \right)\,\frac{d}{dr}-(u+v)(u+v+2)+\frac{(u-v)^2}{r^2}  \right] \,W_{u,v}(r,\phi)\,,
\] 
so that for any linear combination $f(r,\phi)$ of the Zernike functions $W_{u,v}(r,\phi)$, we have that
\[
\label{60}
D_R^2\,f(r,\phi)
\ds =\frac1{1-R^2}\;\left[ \left( 3R-\frac1R  \right)\,D_R-(U+V)(U+V+2)+\frac1{R^2}\,(U-V)^2  \right]\,f(r,\phi)\,.
\]
\end{widetext}
This equation  gives us a formal relation between $D_R^2$ and $D_R$. This is quite interesting, since this allows us to write any operator of the form (\ref{58}) as a first order differential operator. As an example, we see that $A_+^2$ can be written as
\[\label{61}
A_+^2=\frac{e^{2i\Phi}}{4}\,
\left[ h(U,V,R)\,D_R+k(U,V,R)\right]\,,
\]
where $h(U,V,R)$ and $k(U,V,R)$ are given functions of the operators $U$, $V$ and $R$, and the use of the Zernike equation allows to show a linear dependence of  $A_+^2$ on $D_R$. 
This result has an interesting consequence related with the fact  that each element  of the six dimensional group  $SU(1,1)\otimes SU(1,1)$ can be written as a direct product: 
$g(\mathbf a,\mathbf b)=g_A (\mathbf a)* g_B(\mathbf b)$ with
\be\begin{array}{l}\label{62}
g_A(\mathbf a)=e^{i(a_1(A_++A_-)+ia_2(A_+-A_-)+a_3\,A_3)}\,,\\[2ex]
g_B(\mathbf b)=e^{i(b_1(B_++B_-)+ib_2(B_+-B_-)+a_3\,B_3)} \,,
\end{array}\ee
where $\mathbf a=(a_1,a_2,a_3), \mathbf b=(b_1,b_2,b_3)\in\R^3$.
This shows that if $g\in SU(1,1)\otimes SU(1,1)$, then, $g\in {\rm UEA}[su(1,1)\oplus su(1,1)]$ and, therefore, each  $g(\mathbf a,\mathbf b)$ as above may be written as a differential operator of first order in $D_R$. In conclusion, we have the following result
\[
g(\mathbf a,\mathbf b)=h_{\mathbf a,\mathbf b}(U,V,R,\Phi)\,D_R+k_{\mathbf a,\mathbf b}(U,V,R,\Phi)\,,
\]
where $h_{\mathbf a,\mathbf b}(U,V,R,\Phi)$ and $k_{\mathbf a,\mathbf b}(U,V,R,\Phi)$ are functions on the given arguments. For practical purposes, one truncates the series that yield to the exponentials (\ref{62}) so as to obtain a simpler although sufficient approximation.  

\section{Potential applications: Soft adaptive optics}\label{softoptics}

As Zernike functions have played a role in optical image processing, we have considered interesting to add a short section on possible applications which may even open a way for future research. This is an operator formalism on the space of Zernike functions intended to be applicable to optical image processing or adaptive optics \cite{tyson2015}.. This is an algebraic procedure that we call {\it soft adaptive optics}. As a tool for image processing, we consider that soft adaptive optics, as any other manipulator of images, can improve other widely used methods. We believe that it could be an interesting tool to enhance the quality of images. In principle, it may offer some advantages  as is not a complicated procedure and the original image is saved, so that it may undergo further manipulations.

Thus in what follows, we qualitatively sketch the applications to soft adaptive optics to the previous formalism.  An elaborated example would have been quite interesting to illustrate the method. However, we have realised that the construction of such an example  is far from trivial and could be the subject of another article. In any case, it goes beyond the scope of the present paper. Nevertheless, we add some figures at the end taken from numerical experiments. This is given in Appendix D.

Let us consider a real function $f(r,\phi)\in L^2(\mathcal D)$. The images on the unit disk $\mathcal D$ are described by 
$\big|f(r,\phi)\big|^2$. Obviously, $\big|f(r,\phi)\big|\in L^2(\mathcal D)$, so that,  \eqref{9}
\[
\big|f(r,\phi)\big|=\sum_{u,v=0}^\infty {\mathfrak f}_{u,v}\,W_{u,v}(r,\phi)\,.
\]

Relation \eqref{10} allows us to obtain the components ${\mathfrak f}_{u,v}$ in terms of the basis $\{W_{u,v}(r,\phi)\}$ as
\[\label{64}
{\mathfrak f}_{u,v}= \int_0^{2\pi}d\phi\int_0^1 dr\, r\, W^*_{u,v}(r,\phi)\, |f(r,\phi)|\,.
\]
Note that the properties of Zernike functions, in particular \eqref{6}, show that ${\mathfrak f}_{u,v}={\mathfrak f}^*_{v,u}$. In practical numerical calculations, we need truncation of the series spanning  the function $|f(r,\phi)|$ in terms of the coefficients ${\mathfrak f}_{u,v}$ and the Zernike functions, so that
\begin{equation}\label{65}
|f(r,\phi)| \approx \sum_{u,v=0}^{u_M,v_M} {\mathfrak f}_{u,v}\, W_{u,v}(r,\phi)\,,
\end{equation}
where $u_M$ and $v_M$ denote the maximum values of $u$ and $v$ in the sum \eqref{9}, respectively, and are related to the digitalisation of the image. 

Then, any operator $\mathcal O$  performing a transformation from the initial image $\big|f(r,\phi)\big|$ to a final image 
$\big|g(r,\phi)\big|$, i.e.
\[
\mathcal O\,:\, \big| f(r,\phi)\big|\;\longrightarrow\; \big| g(r,\phi)\big|\,,
\]
 has the form of the sum \eqref{58}, where we have to replace the operators $A$ and $B$ acting on kets $|a,b\rangle$ by $\mathcal A$ and $\mathcal B$ acting on the Zernike functions. We have to take into account that  the term in $\overline\alpha,\overline\beta$ obeys the following approximate identity:
 \begin{widetext}
 \be\label{66}
\mathcal O_{\overline\alpha,\overline\beta}\, \big|f(r,\phi)\big| = \sum_{u,v=0}^{u_M,v_M} {\mathfrak f}_{u,v}\,  c_{\overline\alpha,\overline\beta}\, {\mathcal A}_+^{\alpha_1}\,{\mathcal A}_3^{\alpha_2}\,{\mathcal A}_-^{\alpha_3}\,{\mathcal B}_+^{\beta_1}\,{\mathcal B}_3^{\beta_2}\,{\mathcal B}_-^{\beta_3}\, W_{u,v}(r,\phi)\,.
\end{equation}
In order to calculate the terms in this sum \eqref{66}, we need to use the identities \eqref{44} and \eqref{51}. This gives a result of the following form
\be\label{67}
{\mathcal A}_+^{\alpha_1}\,{\mathcal A}_3^{\alpha_2}\,{\mathcal A}_-^{\alpha_3}\,{\mathcal B}_+^{\beta_1}\,{\mathcal B}_3^{\beta_2}\,{\mathcal B}_-^{\beta_3}\, W_{u,v}(r,\phi) = g_{u+\alpha_1-\alpha_3, v+\beta_1-\beta_3}\, W_{u+\alpha_1-\alpha_3, v+\beta_1-\beta_3} (r,\phi)\,,
\ee
where the coefficients $g_{k,l}$ are given by 
\be\label{670}
g_{u+\alpha_1-\alpha_3, v+\beta_1-\beta_3}=(u-\alpha_3+1)^{(\alpha_1)}\,(u-\alpha_3+1/2)^{\alpha_2}\,(u)_{\alpha_3} 
\, (v-\beta_3+1)^{(\beta_1)}\,(v-\beta_3+1/2)^{\beta_2}\,(v)_{\beta_3}\,,
\ee
 \end{widetext}
with 
\[
(x)_n:= x (x-1) (x-2)\cdots  (x-n+1)=\frac{\Gamma(x+1)}{\Gamma(x-n+1)}
\]
the falling factorial and 
\[
x^{(n)}:= x (x+1) (x+2)\cdots  (x+n-1)=\frac{\Gamma(x+n)}{\Gamma(x)}
\]
the raising factorial or  Pochhammer symbol.
 This obviously gives
\[ \begin{array}{l}\label{68}
\mathcal O_{\overline\alpha,\overline\beta}\, \big|f(r,\phi)\big|
\ds = \sum_{u,v=0}^{u_M,v_M} {\mathfrak f}_{u,v} \, c_{\overline\alpha,\overline\beta}\,\, g_{u+\alpha_1-\alpha_3, v+\beta_1-\beta_3}\, 
 \\[0.3cm] 
 \hskip 3cm \ds\times  W_{u+\alpha_1-\alpha_3, v+\beta_1-\beta_3} (r,\phi)\,,
 \end{array}\]
a result that provides the final expression of the object image  as 
\[
\big|g(r,\phi)\big|=\mathcal O\, \big|f(r,\phi)\big|\,.
\] 

In Appendix D, we add an illustration of the proposed procedure.

\section{Conclusions}

We have revisited some properties of the Zernike polynomials and their connections with the symmetry group $SU(1,1)\otimes SU(1,1)$. The introduction of the W-Zernike functions $W_{u,v}(r,\phi)$ in \eqref{5} is important in order to find in a very natural way the ladder operators that close the Lie algebra $su(1,1)\oplus su(1,1)$. 

The Zernike functions can be seen as transition matrices between continuous and discrete bases on the unit disk 
$\mathcal D$, 
$\{|r,\phi\rangle\}_{(r,\phi)\in\mathcal D}$ and $\{|u,v\rangle\}_{u,v\in \N}$ respectively.

It is well known that 
discrete and continuous basis are quite often used in quantum physics although the continuous basis is does not exist in the Hilbert space, which is often considered as the usual framework of quantum mechanics. This is why  the  formalism  of rigged Hilbert spaces is needed, where both types of bases acquire full meaning. In this context a continuous basis is a set of functionals over a space of test vectors. Some formal and useful relations between both kinds of bases are presented.

The elements of the  Lie algebra $su(1,1)\oplus su(1,1)$ are unbounded as operators on  Hilbert spaces.  However, as operators on rigged Hilbert space all these unbounded operators  become continuous. The same happens with the elements of the UEA[$su(1,1)\oplus su(1,1)$] and, obviously,  with the elements of the group $SU(1,1)\otimes SU(1,1)$. 

These properties are interesting in possible applications to  soft adaptive optics where the original image can be transformed to a new one by means of continuous operators of UEA[$su(1,1)\oplus su(1,1)$].


\section*{Acknowledgments}
This research is supported in part by the Ministerio de Econom\'ia y Competitividad of Spain  under grant  MTM2014-57129-C2-1-P and the Junta de Castilla y Le\'on (Projects VA057U16, VA137G18 and BU229P18).


\section*{Appendix A: Zernike polynomials for $| r|\leq 1$}

The Zernike polynomials $R^m_n(r)$ can be enlarged for negative valued of r, i.e., $r\in [-1,1]$, Hence 
\be\label{Zgeneral}
R^m_n(-r)=(-1)^n\, R^m_n(r)\,.
\ee
In this case the new orthogonality and completeness relations are now
\be \begin{array}{l}\label{3a}
\ds \int_{-1}^1 R^m_n(r)\,(n+1)\,R^m_{n'}(r)\,|r|\, dr= \delta_{n,n'}
\,,\\[0.4cm]
\ds \sum_{n=|m|}^\infty  R^m_n(r)\,R^m_n(r')\,(n+1)=\delta(r^2-r'^2)\\[0.4cm]
\hskip1.25cm \ds =\frac{1}{2|r|}\,\left(\delta(r+r')+\delta(r-r')\right)\,.
\end{array}\ee 

The symmetries of the Zernike radial polynomials  determine the Lie group $SU(1,1)\otimes SU(1,1)$ \cite{WUN,Celeghini}.
Its Lie infinitesimal generators in the representation $R^{|u-v|}_{u+v}(r)$ have the following explicit form  valid
 for $|r|\leq 1$
 \[\begin{array}{l}
{\cal A}_\pm  :=  \frac{1}{2} \mp (1-r^2) D_r +r(U+V+1\pm 1) \\[0.3cm]
\hskip2.5cm +\frac{1}{r} (U-V)]\,,
\\[0.4cm]
{\cal B}_\pm  :=  \frac{1}{2} [\mp (1-r^2) D_r +r(U+V+1\pm 1) \\[0.3cm]
\hskip2.5cm  -\frac{1}{r} (U-V)]\,.
\end{array}\]
Their action on $R_{u+v}^{|u-v|}(r)$  is 
\[\begin{array}{l}
{\cal A}_\pm  R_{u+v}^{|u-v|}(r) =  (u+\frac 12\pm \frac 12) \,R_{(u\pm 1)+v}^{|(u\pm 1)-v|} (r)\,,
\\[0.4cm]
{\cal B}_\pm  R_{u+v}^{|u-v|}(r) =  (v+\frac 12\pm \frac 12)\, R_{u+(v\pm 1)}^{|u-(v\pm 1)|} (r)\,.
\end{array}\]
\section*{Appendix B: Proof of relation (\ref{23})}

Let us consider the Zernike polynomial $R_n^m(r)$ and since the symmetry property $R^m_n(r)=R^{-m}_n(r)$ we can considerer $m\geq 0$ without loss of generality. This is a polynomial for which monomials $r^k$ are either even or odd. In the first case, $n$ and $m$ are both even and both odd in the second case. The first term is proportional to $r^{n}$ and the last one to $r^{m}$, so that a typical Zernike polynomial has the form
\[ \begin{array}{l}\label{69}
R_n^m(r)=
a_{n}\,r^{n}+a_{n-2}\,r^{n-2}+\dots+a_{m}\, r^{m}\,,
\end{array}\]
where the $a_i$ are real numbers. Multiplying  $R_n^m(r)$ by $r$ we have
\[\begin{array}{l}\label{70}
r\,R_n^m(r)=a_{n}\,r^{n+1}+a_{n-2}\,r^{n-1}+\dots+a_{m}\, r^{m+1}\,.
\end{array}\]
Since the Zernike polynomials $R_{n+1}^{m+1}(r)$, $R_{n-1}^{m+1}(r)$, $R_{n-3}^{m+1}(r),\dots,R_{n-m+1}^{m+1}(r)$ are linearly independent polynomials of degree $n+1$, $n-1$, \dots, $m+1$, respectively, we have that
\[ \begin{array}{l} \label{71}
a_{n}\,r^{n+1}+a_{n-2}\,r^{n-1}+\dots+a_{m}\, r^{m+1}\\[0.3cm]
\hskip1cm  = b_{n+1}\,R_{n+1}^{m+1}(r)+b_{n-1}\,R_{n-1}^{m+1}(r) \\[0.3cm] 
\hskip3cm +\dots + b_{m+1} \,R_{m+1}^{m+1}(r)\,,
\end{array} \]
where the coefficientes $b_{k}$ are real numbers. In conclusion,
\[\begin{array}{l}\label{72}
r\,R_n^m(r)=  b_{n+1}\,R_{n+1}^{m+1}(r)+b_{n-1}\,R_{n-1}^{m+1}(r)\\[0.3cm]
\hskip3cm +\dots + b_{m+1} \,R_{m+1}^{m+1}(r)\,,
\end{array}\]

However we can refine prove the previous relation  between Zernike polynomials. So we can establish the following 
\medskip

\noindent
{\bf Property 1.-} {\sl  Any Zernike polynomial $R_n^m(r)$ such that $n\geq 1$ verifies the  relation
\be\label{Zpropiedad}
r\,R_n^m(r)=a_n^m\,R_{n+1}^{m+1}(r)+b_n^m\,R_{n-1}^{m+1}(r)
\ee
where
\[
a_n^m=\frac{n+m+2}{2(n+1)}\,,\qquad b_n^m=\frac{n-m}{2(n+1)}\,. 
\]}

\noindent
{\bf Proof.-} Effectively, let us start with this explicit formula of the Zernike polynomials $R_n^m$
\eqref{Zformula}
Now from \eqref{Zformula} the l.h.s. of \eqref{Zpropiedad} can be written as

\be\begin{array}{l}\label{Zlhs}
r\,R_n^m(r)=\ds \sum_{k=0}^{\frac{n-m}{2}} (-1)^k\,\left( \begin{array}{c} n-k\\ k\end{array}\right)\\[0.3cm]
\hskip2cm \ds
\times \left( \begin{array}{c} n-2k\\ \frac{n-m}{2}- k\end{array}\right)\,r^{n+1-2k}
\end{array}\ee
Also the  r.h.s. of \eqref{Zpropiedad} is equal to
\begin{widetext}
\[\label{Zrhs}
 a\,\sum_{k=0}^{\frac{n-m}{2}} (-1)^k\,\left( \begin{array}{c} n+1-k\\ k\end{array}\right)\,\left( \begin{array}{c} n+1-2k\\ \frac{n-m}{2}- k\end{array}\right)\,r^{n-2k}
\ds+\, b\,\sum_{k=0}^{\frac{n-m-2}{2}} (-1)^k\,\left( \begin{array}{c} n-1-k\\ k\end{array}\right)\,\left( \begin{array}{c} n-1-2k\\ \frac{n-m-2}{2}- k\end{array}\right)\,r^{n-1-2k}
\]
that we can rewrite as
\be\label{Zrhs}
\ds \sum_{k=0}^{\frac{n-m}{2}} (-1)^k\,\left[a\,\left( \begin{array}{c} n+1-k\\ k\end{array}\right)\,
\left( \begin{array}{c} n+1-2k\\ \frac{n-m}{2}- k\end{array}\right)\right. 
\ds\left.-\, b\,(1-\delta_{0,k})\,\left( \begin{array}{c} n-k\\ k-1\end{array}\right)\,\left( \begin{array}{c} n+1-2k\\ \frac{n-m}{2}- k\end{array}\right)\right]\,r^{n+1-2k}
\ee
From \eqref{Zlhs} and \eqref{Zrhs} we obtain the following relations for the coefficients of the powers of $r^{n+1-2k}$ for $k=0,1,\cdots, (n-m)/2$
\[\label{Zcoefficients}\begin{array}{l}
 \left( \begin{array}{c} n-k\\ k\end{array}\right)\,
\left( \begin{array}{c} n-2k\\ \frac{n-m}{2}- k\end{array}\right)
=
a\,\left( \begin{array}{c} n+1-k\\ k\end{array}\right)\,
\left( \begin{array}{c} n+1-2k\\ \frac{n-m}{2}- k\end{array}\right)
\ds -\, b\,(1-\delta_{0,k})\,\left( \begin{array}{c} n-k\\ k-1\end{array}\right)\,\left( \begin{array}{c} n+1-2k\\ \frac{n-m}{2}- k\end{array}\right)
\end{array}\]
\end{widetext}
From $k=0$ the previous expression becomes
\[
\left( \begin{array}{c} n\\ \frac{n-m}{2}\end{array}\right)
=a\,\left( \begin{array}{c} n+1\\ \frac{n-m}{2}\end{array}\right)
\]
So from the definition of the binomial coefficients we obtain that
\be\label{a}
a=\frac{n+m+2}{2(n+1)}
\ee
and for $k\geq 1$  we get
\begin{widetext}
\[\begin{array}{l}
 \left( \begin{array}{c} n-k\\ k\end{array}\right)\,
\left( \begin{array}{c} n-2k\\ \frac{n-m}{2}- k\end{array}\right)
=
a\,\left( \begin{array}{c} n+1-k\\ k\end{array}\right)\,
\left( \begin{array}{c} n+1-2k\\ \frac{n-m}{2}- k\end{array}\right)
 -\, b\,\left( \begin{array}{c} n-k\\ k-1\end{array}\right)\,\left( \begin{array}{c} n+1-2k\\ \frac{n-m}{2}- k\end{array}\right)
\end{array}\]
\end{widetext}
that developing the binomial coefficients in terms of factorial we get
that
\[
1=a\,\frac{2(n+1-k)}{n+m+2-2k}-b\,\frac{2 k}{n+m+2-2k}
\]
And now from \eqref{a} we get that
\[
b=\frac{n-m}{2(n+1)}\,,
\]
which is independent of $k$.
\hfill $\blacksquare$\medskip

As a corolary we have the following property of the Zernike functions $W_{u,v}(r,\phi )$
\medskip

\noindent
{\bf Property 2.-} 
Any Zernike function $W_{u,v}(r,\phi)$ such that $n\geq 1$ verifies the following relation
\[\label{Wpropiedad}
r\,e^{i\phi}\,W_{u,v}(r,\phi)=\a_u^v\,W_{u+1,v}(r,\phi)+\b_u^v\,W_{u,v-1}(r,\phi)\,,
\]
where 
\[\begin{array}{l}\label{Wcoefficients}
\ds \a_u^v=\frac{u+1}{\sqrt{(u+v+1)(u+v+2)}}\,,\\[0.3cm]
\ds \b_u^v=\frac{v}{\sqrt{(u+v)(u+v+1)}}\,.
\end{array}\]

\noindent
{\bf Proof.-} The proof is trivial taking into account the definition \eqref{5} of the $W_{u,v}(r,\phi )$ as well as the relations \eqref{4} between the parameters $(u,v)$ and $(n,m)$.\hfill $\blacksquare$\medskip

After the definition of $\Phi$, this proves the stability of $\Phi$ under the action of $P$. Note that we have not proved the continuity of $P$ on $\Phi$, neither some topological properties of $\Phi$ with respect to the topology inherited from $\Phi_1$. This is not strictly necessary for our purposes. 

\section*{Appendix C: Another topology for the space of Zernike functions}

Along with the space $\Psi_1$ of functions $f(r,\theta)$ of $L^2(\mathcal D)$ \eqref{9} 
such that
\[
\sum_{u,v=0}^\infty \big|f_{u,v}\big|^2\,(u+v+1)^{2p}<\infty\,,\;\;\; \forall p\in N \,,\]
we consider another one that we denote here as $\Psi$. This is the space of functions $f(r,\theta) \in L^2(\mathcal D)$ \eqref{9}  verifying
\begin{equation*}\label{a1}
\sum_{u,v}^\infty \big| f_{u,v}\big|\,(u+v+1)^q<\infty\,,\;\forall q\in \N\,.
\end{equation*}
We endow $\Psi$ with the  set of norms $\big|\big|-\big|\big|_{1,q}$  
\begin{equation*}\label{a2}
\big|\big|f(r,\phi)\big|\big|_{1,q}:= \sum_{u,v}^\infty \big| f_{u,v}\big|\,(u+v+1)^q\,,\end{equation*}
with $q= 0,1,2,\dots$, so that $\Psi$ has the structure of countably normed space and, hence, metrizable. 
\medskip

This space has the following properties:
\begin{enumerate}
\item
 {It is dense in $L^2(\mathcal D)$, since it contains all the basis elements $W_{u,v}(r,\phi)$. }

\item {The series
\begin{equation}\label{a3}
f(r,\phi)= \sum_{u,v=0}^\infty f_{u,v}\, W_{u,v}(r,\phi)
\end{equation}
converges absolutely and uniformly and hence point-wise. 

The proof is the following: since the functions $W_{u,v}(r,\phi)$ have the  upper bound \eqref{upperbound}, i.e.
$\big|W_{u,v}(r,\phi)\big|\le \sqrt{(u+v+1)(\pi)}\,,$
then,
\begin{equation*}\begin{array}{l}\label{a5}
\ds\sum_{u,v=0}^\infty \big|f_{u,v}\big|\cdot \big|W_{u,v}(r,\phi)\big| \\[0.4cm]
\hskip1cm \ds\le \sum_{u,v=0}^\infty \big|f_{u,v}\big|\,
\sqrt{\frac{u+v+1}{\pi}}\\[0.4cm]
\hskip1cm \ds \le \frac 1{\sqrt\pi} \sum_{u,v=0}^\infty \big|f_{u,v}\big|\,(u+v+1)<\infty\,.
\end{array}\end{equation*}
Then, the Weiersstrass M-Theorem guarantees the absolute and uniform convergence of the series. 
}

\item  {Observe that for all absolutely convergent series $\sum_n a_n$, we have that
\begin{equation*}\label{a6}
\sqrt{\sum_n\big|a_n\big|^2} \le \sum_n \big|a_n\big|\,.
\end{equation*}
This shows that, if $\big|\big|-\big|\big|_r$ is the norm defined in (\ref{21}), we have that
\begin{equation*}\begin{array}{ll}\label{a7}
\big|\big|f\big|\big|_r &=\ds \sqrt{\sum_{u,v}^\infty \big|f_{u,v}\big|^2\,(u+v+1)^{2r}} \\[0.5cm]
& \ds\le \sum_{u,v}^\infty \big|f_{u,v}\big|\,(u+v+1)^r=: p_r(f)\,,
\end{array}\end{equation*}
which shows that $\Psi\subset \Psi_1$ and also that the canonical injection 
\[\begin{array}{lc}
\Psi\,\stackrel{i}\longrightarrow\, &\Psi_1\\[0.3cm]
 f \,\longrightarrow\,&i(f)=f
 \end{array}\] 
is continuous. This implies that the canonical injection $i:\Psi\longmapsto L^2(\mathcal D)$ is also continuous, so that $\Psi\subset L^2(\mathcal D)\subset \Psi^\times$ is a rigged Hilbert space. }

\item {The operators $\mathcal U, \mathcal V, \mathcal A_\pm,\mathcal B_\pm$ are continuous on $\Psi$. 

The proof is straightforward.  It is also important to show that the operator $P$, as defined in \eqref{23} and (\ref{Wcoeff}) is invariant and continuous on $\Phi$. The proof is very simple. For any $f(r,\phi)\in\Psi$ as in \eqref{a3}, we have that  
\begin{equation*}\begin{array}{l}\label{a8}
\ds P\sum_{u,v=0}^\infty f_{u,v} W_{u,v}(r,\phi) = \sum_{u,v=0}^\infty \left(\a_{u-1}^vf_{u-1,v}\right.
 \\[0.3cm]
 \ds \hskip2cm \left.+\b_u^{v+1}f_{u,v+1}\right)  W_{u,v}(r,\phi)\,,
\end{array}\end{equation*}
where we take $f_{-1,v}=0$. Since $0\le \a_u^v\le 1$ and $0\le \b_u^v\le 1$, we get that
\begin{equation*}\begin{array}{l}\label{a9}
\ds\bigg|\bigg|P\sum_{u,v=0}^\infty f_{u,v}\, W_{u,v}(r,\phi)\bigg|\bigg|_{1,r} 
\\[0.4cm]\ds 
= \sum_{u,v=0}^\infty \big|\a_{u-1}^v\,f_{u-1,v}+\b_{u}^{v+1}\,f_{u,v+1}\big|\,(u+v+1)^r \\[0.4cm] 
\hskip0.75cm \ds\le \sum_{u,v=0}^\infty \big|f_{u-1,v}\big|\,(u+v+1)^r  \\[0.4cm]
\hskip1.95cm \ds+ \sum_{u,v=0}^\infty \big|f_{u,v+1}\big|\,(u+v+1)^r\,.
\end{array}\end{equation*}
The first term of the last inequaliity in the previous expression  since $f_{-1,0}=0$ can be rewritten as
\begin{equation*}\begin{array}{l}\label{a10}
\ds\sum_{u,v=0}^\infty |f_{u-1,v}|\,(u+v+1)^r 
  \\[0.4cm]  \hskip1.25cm \ds 
  = \sum_{u,v=0}^\infty \big|f_{u,v}\big|\,(u+v+2)^r   \\[0.4cm]  
 \hskip1.25cm \ds\le 2^r\sum_{u,v=0}^\infty \big|f_{u,v}\big|\,(u+v+1)^r    
 \\[0.4cm]  \hskip1.25cm 
 \ds = 2^r \,\bigg|\bigg|\sum_{u,v=0}^\infty f_{u,v}\, W_{u,v}(r,\phi)\bigg|\bigg|_{1,r}\,,
\end{array}\end{equation*}
while the second one gives
\begin{equation*}\begin{array}{l}\label{a11}
\ds\sum_{u,v=0}^\infty |f_{u,v+1}|\,(u+v+1)^r \\[0.4cm]\ds  \hskip1.25cm
\le \sum_{u,v=0}^\infty \big|f_{u,v}\big|\, (u+v)^r\\[0.4cm]\ds  \hskip1.25cm
\le \sum_{u,v=0}^\infty \big|f_{u,v}\big|\, (u+v+1)^r \\[0.4cm]\ds  \hskip1.25cm
= \bigg|\bigg|\sum_{u,v=0}^\infty f_{u,v}\, W_{u,v}(r,\phi)\bigg|\bigg|_{1,r}\,.
\end{array}\end{equation*}
This shows that
\begin{equation*}\begin{array}{l}\label{a12}
\ds\bigg|\bigg|P\sum_{u,v=0}^\infty f_{u,v}\, W_{u,v}(r,\phi)\bigg|\bigg|_{1,r}  \\[0.4cm
]\ds  \hskip1.25cm
\le(2^r+1) \,\bigg|\bigg|\sum_{u,v=0}^\infty f_{u,v}\, W_{u,v}(r,\phi)\bigg|\bigg|_{1,r}\,,
\end{array}
\end{equation*}
which proves our claim. 
}
{From (\ref{32}), we have that
\begin{equation*}
\big|\langle r,\phi|f\rangle\big| =\big|f(r,\phi)\big| \le \sum_{u,v=0}^\infty \big|f_{u,v}\big| = ||f(r,\phi)||_{1,0}\,,
\end{equation*}
so that $\langle r,\phi|$ is a continuous mapping on $\Psi$. 
}
\end{enumerate}
\begin{figure}[t]
\centering
 \subfigure [$\,\zf^{-1}_3(r, \phi)$]{\includegraphics[width=0.20\textwidth]{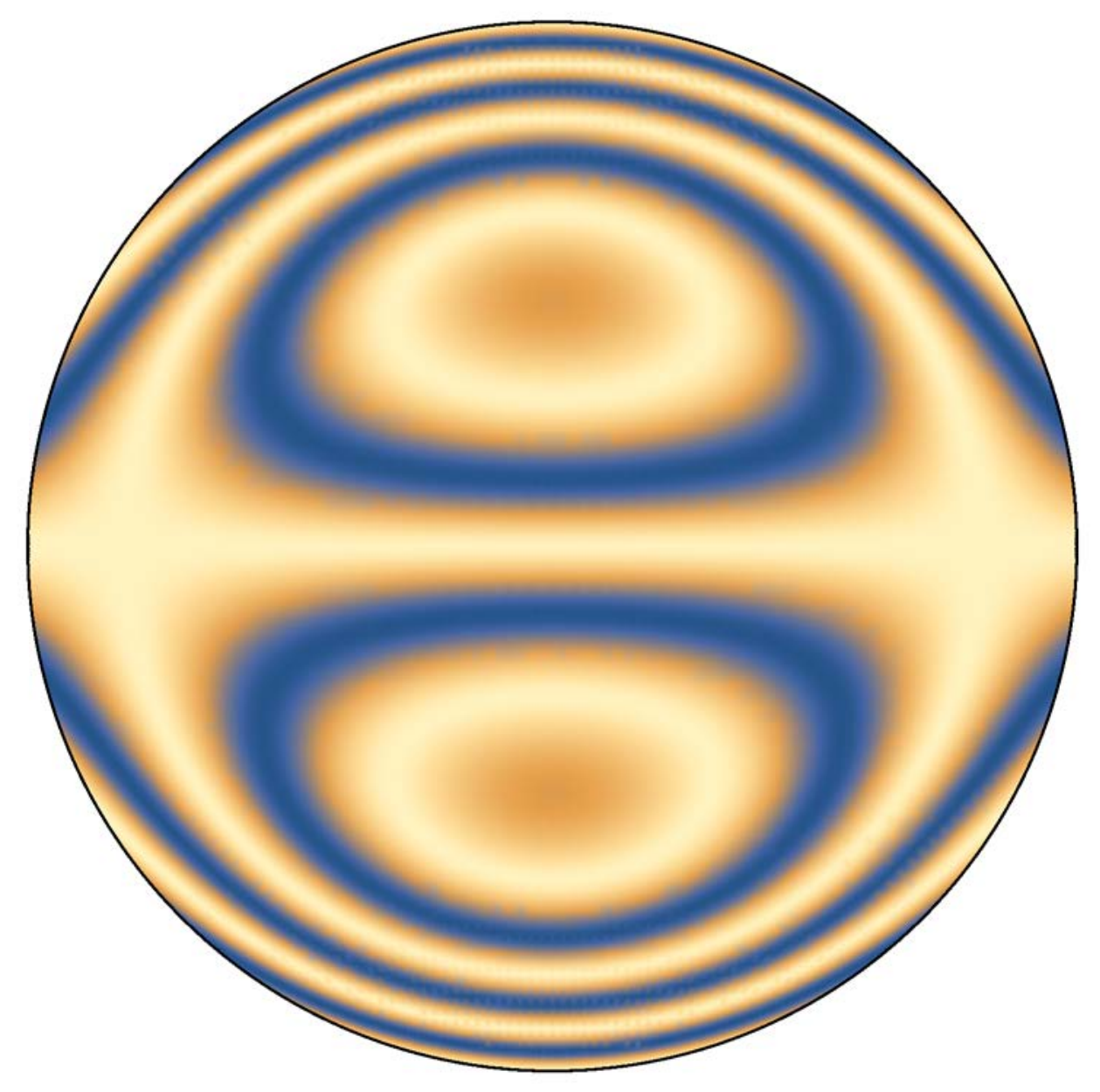}}\qquad  
\subfigure [$\,\text{Im}\, W_{2,1}(r,\phi)$]{\includegraphics[width=0.20\textwidth]{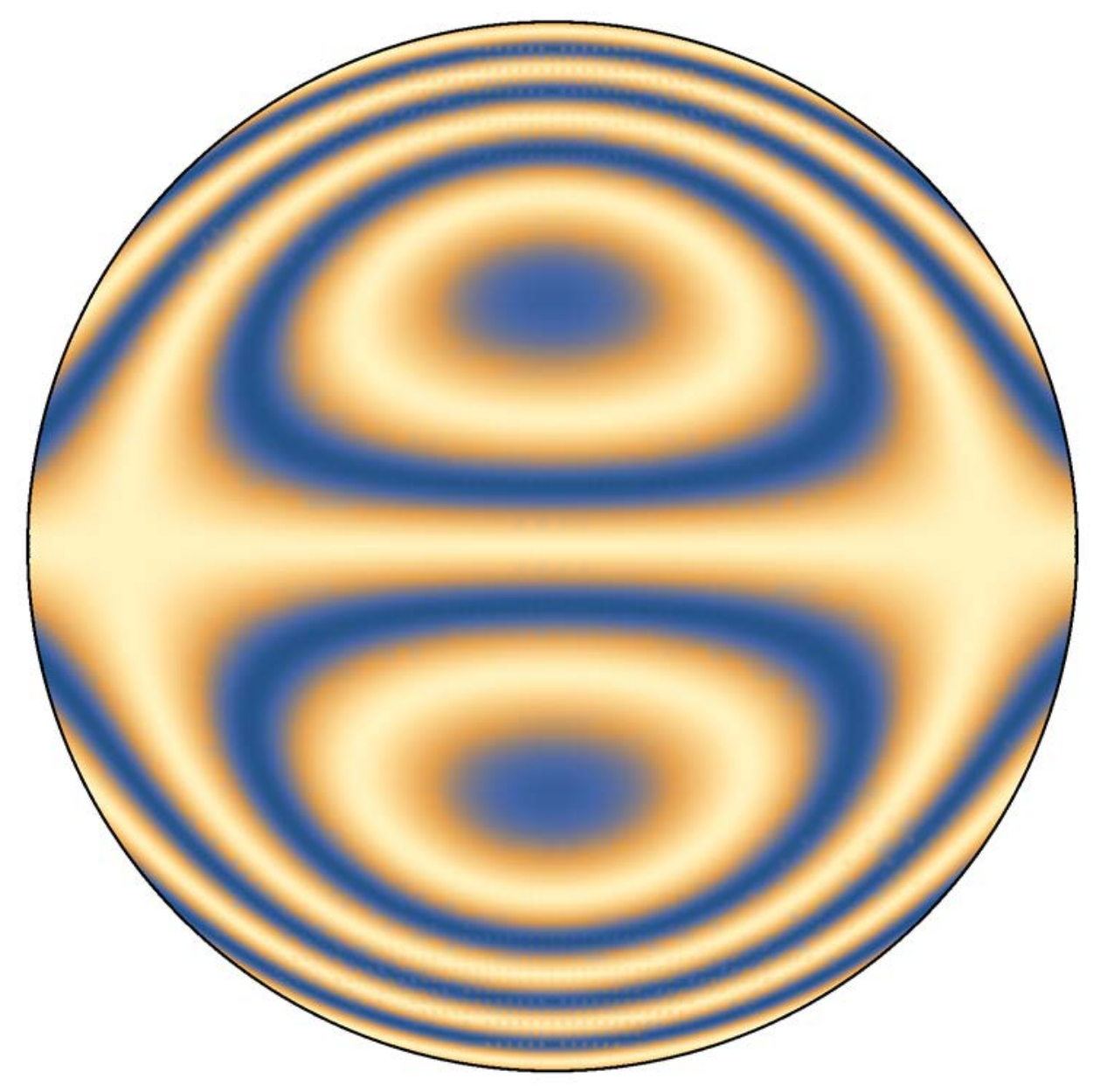}}
\bigskip

 \subfigure [$\,\zf$]{\includegraphics[width=0.20\textwidth]{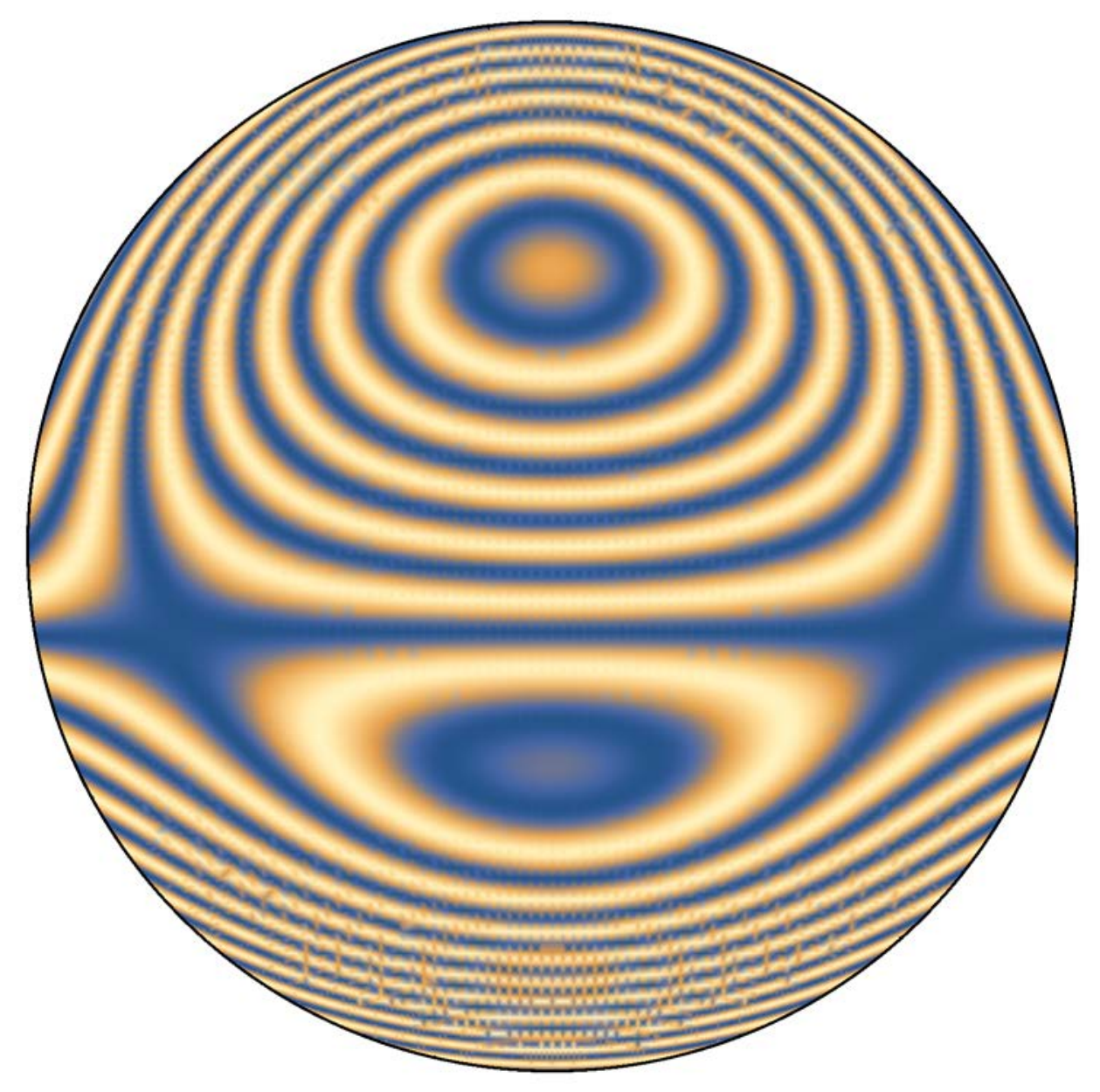}}\qquad  
\subfigure [$\,W$]{\includegraphics[width=0.20\textwidth]{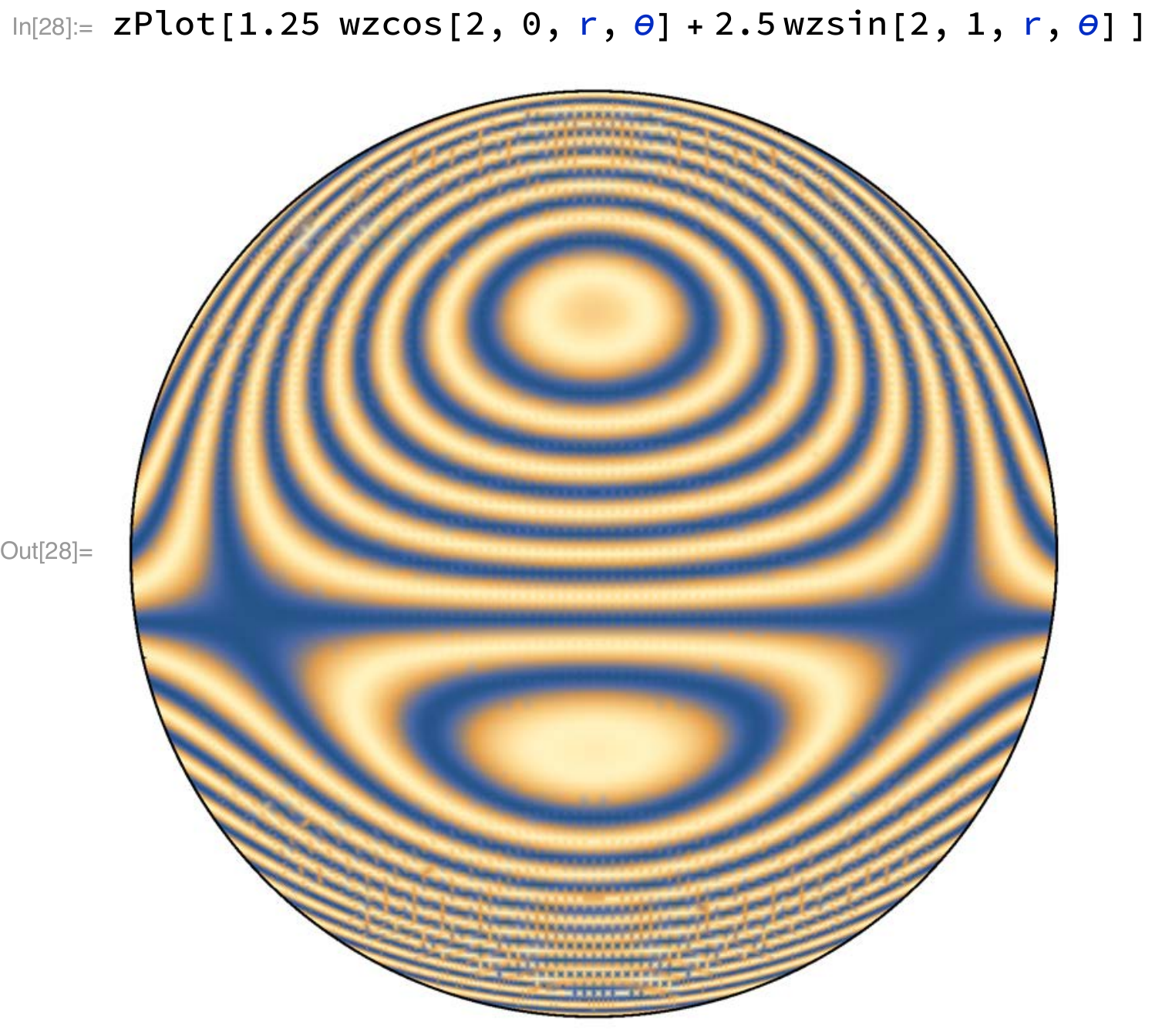}}
\bigskip

 \subfigure [$\, \zf^{16}_{24}$]{\includegraphics[width=0.20\textwidth]{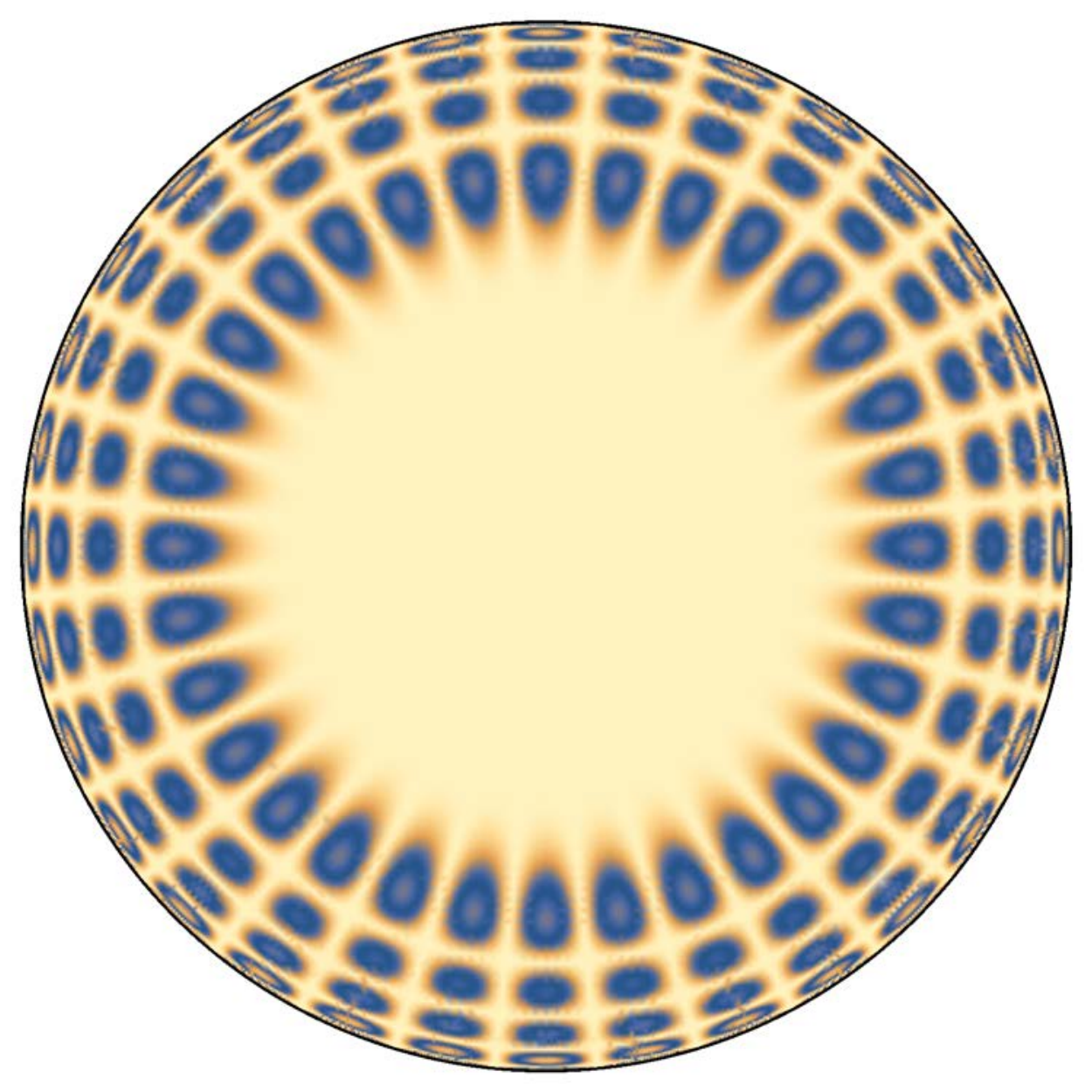}}\qquad  
 \subfigure [$\,\text{Re}\, W_{20,4}$]{
 \includegraphics[width=0.20\textwidth]{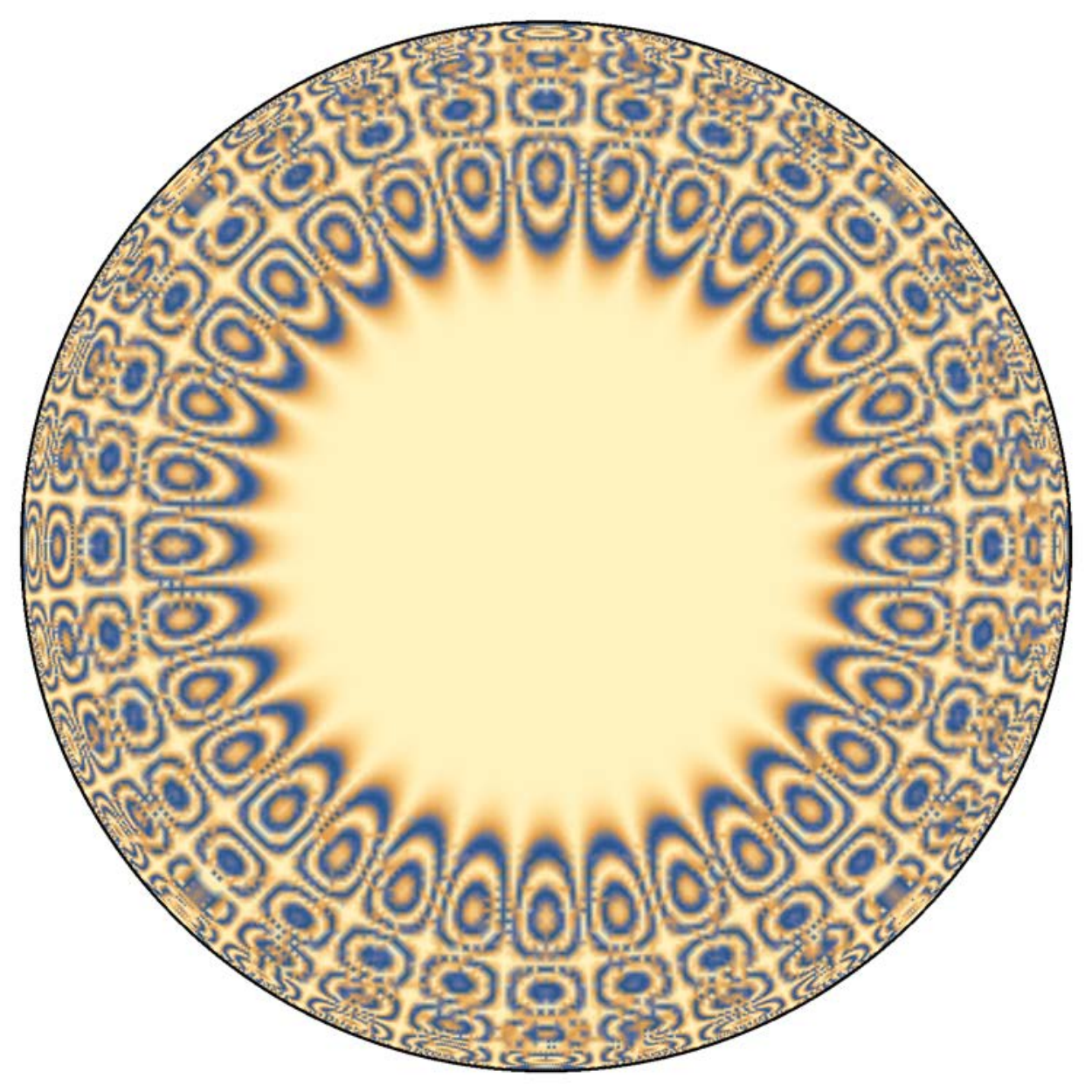}}
\caption{\footnotesize {${\mathcal Z}$-Zernike functions  versus $W$-Zernike functions,
(c) $\zf=1.25\, \zf^{2}_2 +2.5\, \zf^{-1}_3$} and (d) $W=1.25\,\text{Re}\, W_{2,0}+2.5\,\text{Im}\, W_{2,1}\,$
} 
\label{figure1}
\end{figure}

\section*{Appendix D: 
${\mathcal Z}$-Zernike functions  versus $W$-Zernike functions}

In Optics it  is very common the use of the first ${\mathcal Z}$-Zernike functions  which are defined as follows
\be\label{pmzernike}
\begin{array}{lll}
\zf_{n}^{m}(r,\phi)&:=& R^m_n(r)\,\cos (m\phi)\,,\\[0.4cm]
\zf_{n}^{-m}(r,\phi) &:=& R^m_n(r)\,\sin (m\phi)\,,
\end{array}
\ee
with $n,\; m\in \N\,,$.
The other conditions verified by $m$ and $n$ are displayed in expression \eqref{conditions}. Both kinds of ${\mathcal Z}$-Zernike functions \eqref{pmzernike} are included by the use of $e^{i m\phi}$ as we have done in \eqref{zernike}. On the other hand, the $W$-Zernike functions \eqref{5} contain a scale factor and they are well adapted to show the underlying symmetry of the Zernike functions that in the representation given by the  ${\mathcal Z}$-Zernike functions is more difficult to see. Rewritting \eqref{5}
\[\begin{array}{lll}
W_{u,v}(r,\phi)&:= &\ds
\sqrt{\frac{u+v+1}{\pi}}\, {\zf}^{u-v}_{u+v}(r,\phi)\\[0.4cm] &\; =&\ds
\sqrt{\frac{u+v+1}{\pi}}\, R^{|u-v|}_{u+v}(r)\,e^{i(u-v)\phi}\,,
\end{array}\]
we easily can find the relation between the first ${\mathcal Z}$-Zernike functions  versus $W$-Zernike functions.Thus
\begin{widetext}
\[\begin{array}{lll}
W_{0}^{0}(r,\phi)=\sqrt{\frac{1}{ \pi}}\,\zf_{0}^{0}(r,\phi)\,,\;\;
W_{1}^{0}(r,\phi)=\sqrt{\frac{2}{ \pi}}\,\zf_{1}^{1}(r,\phi)\,,\;\;
 W_{1}^{1}(r,\phi)=\sqrt{\frac{3}{ \pi}}\,\zf_{2}^{0}(r,\phi)\,\\[0.4cm]
W_{2}^{0}(r,\phi)=\sqrt{\frac{3}{ \pi}}\,\zf_{2}^{2}(r,\phi)\,,\;\;
 W_{2}^{1}(r,\phi)=\sqrt{\frac{4}{ \pi}}\,\zf_{3}^{1}(r,\phi)\,,\;\;
 W_{2}^{2}(r,\phi)=\sqrt{\frac{5}{ \pi}}\,\zf_{4}^{0}(r,\phi)\,,\\[0.4cm]
 W_{3}^{0}(r,\phi)=\sqrt{\frac{4}{ \pi}}\,\zf_{3}^{3}(r,\phi)\,,\;\;
 W_{3}^{1}(r,\phi)=\sqrt{\frac{5}{ \pi}}\,\zf_{4}^{2}(r,\phi)\,,\;\;
 W_{3}^{2}(r,\phi)=\sqrt{\frac{6}{ \pi}}\,\zf_{5}^{1}(r,\phi)\,,\\[0.4cm]
 W_{3}^{3}(r,\phi)=\sqrt{\frac{7}{ \pi}}\,\zf_{6}^{0}(r,\phi)\,,\;\;
 W_{4}^{1}(r,\phi)=\sqrt{\frac{6}{ \pi}}\,\zf_{5}^{3}(r,\phi)\,,\;\;
 W_{4}^{2}(r,\phi)=\sqrt{\frac{7}{ \pi}}\,\zf_{6}^{2}(r,\phi)\,,\\[0.4cm]
 \dotfill{}
 \end{array}\]

\begin{figure}[t]
\centering
 \subfigure [$\,\text{Re}\, W_{4,1}$]{\includegraphics[width=0.20\textwidth]{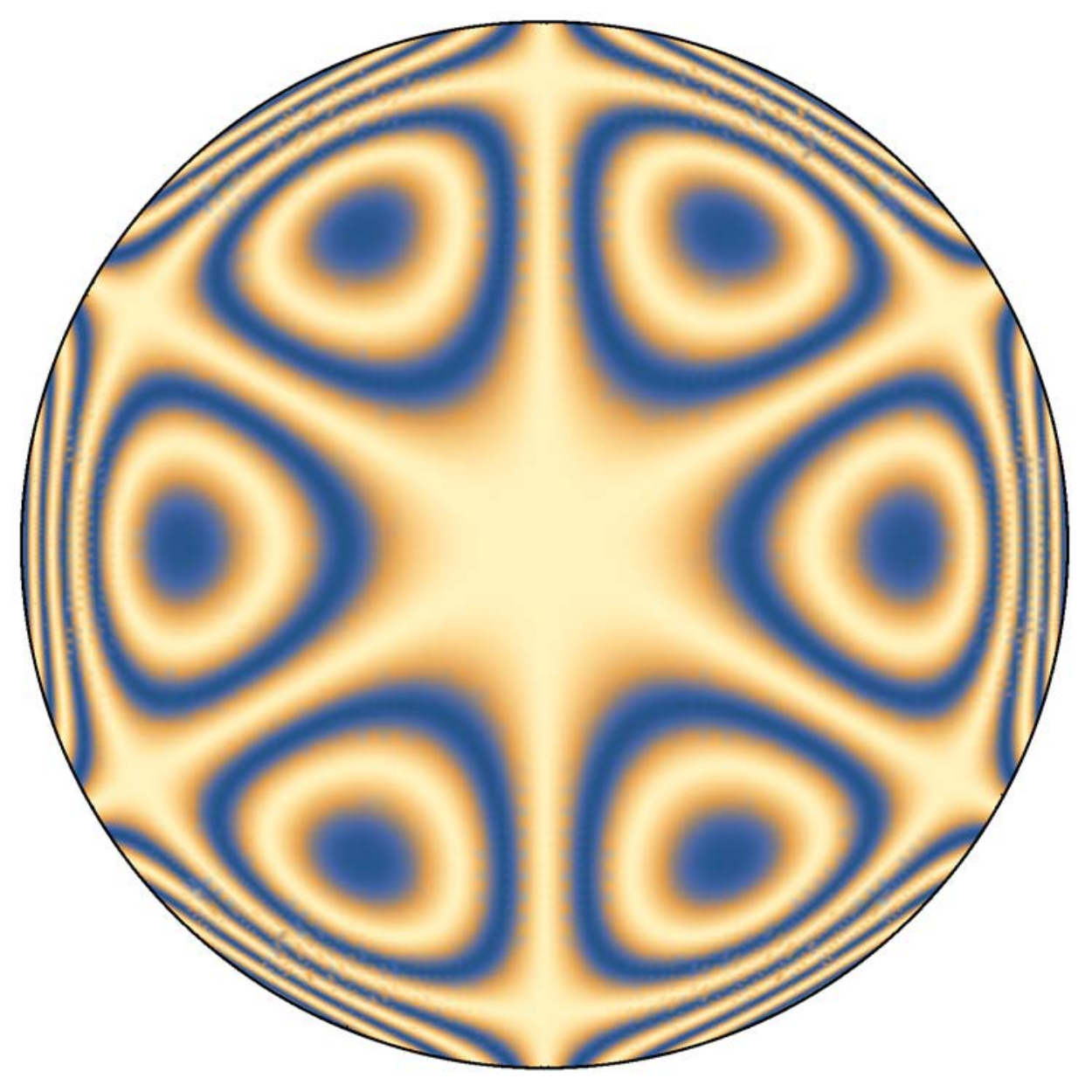}}\quad  
 \subfigure [$\,\text{Re}\,\mathcal O_{\overline\alpha,\overline\beta} W_{4,1}$]{\includegraphics[width=0.20\textwidth]{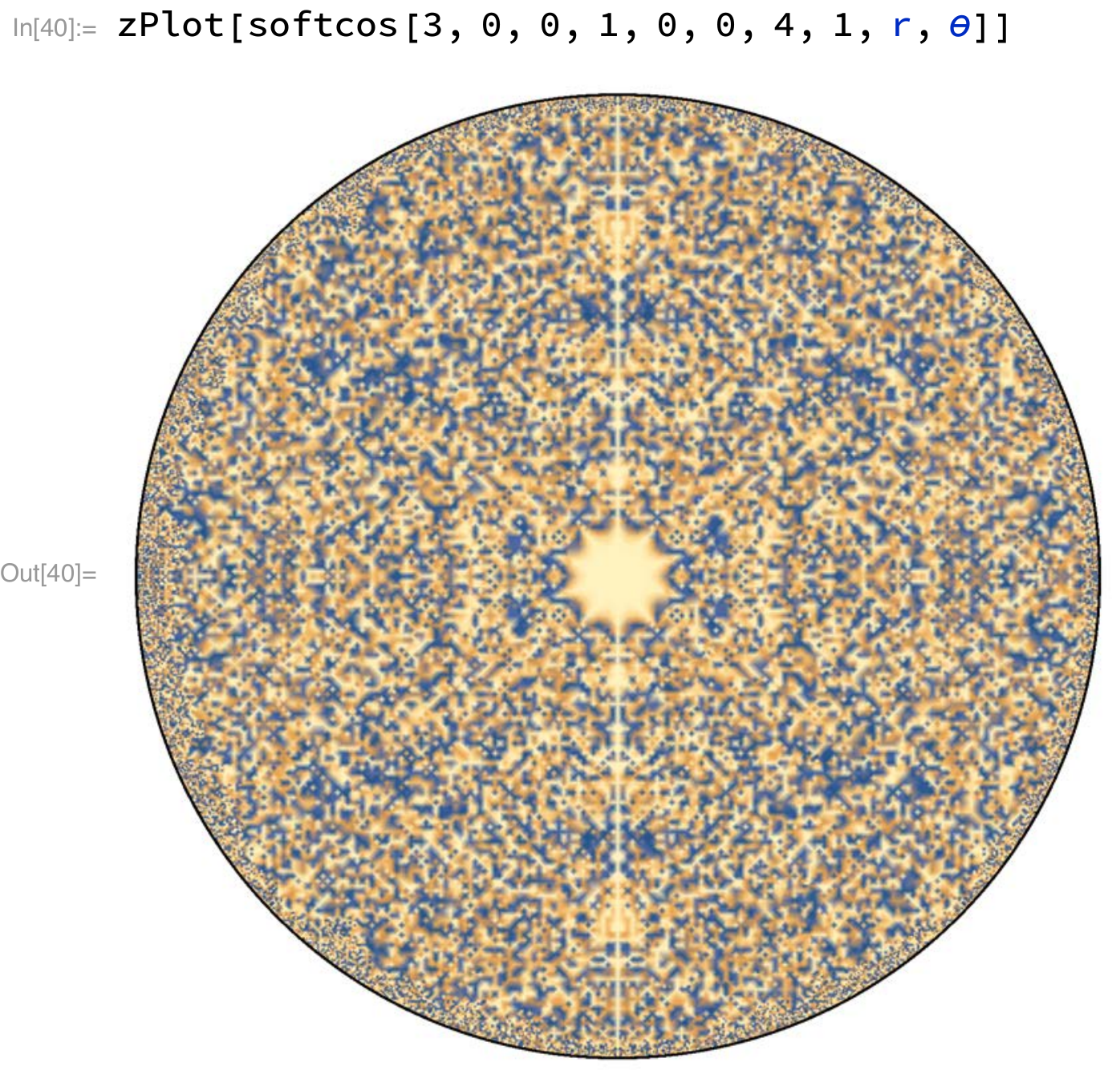}}\quad
 \subfigure [$\,\text{Re}\, W_{7,2}\,$]{\includegraphics[width=0.20\textwidth]{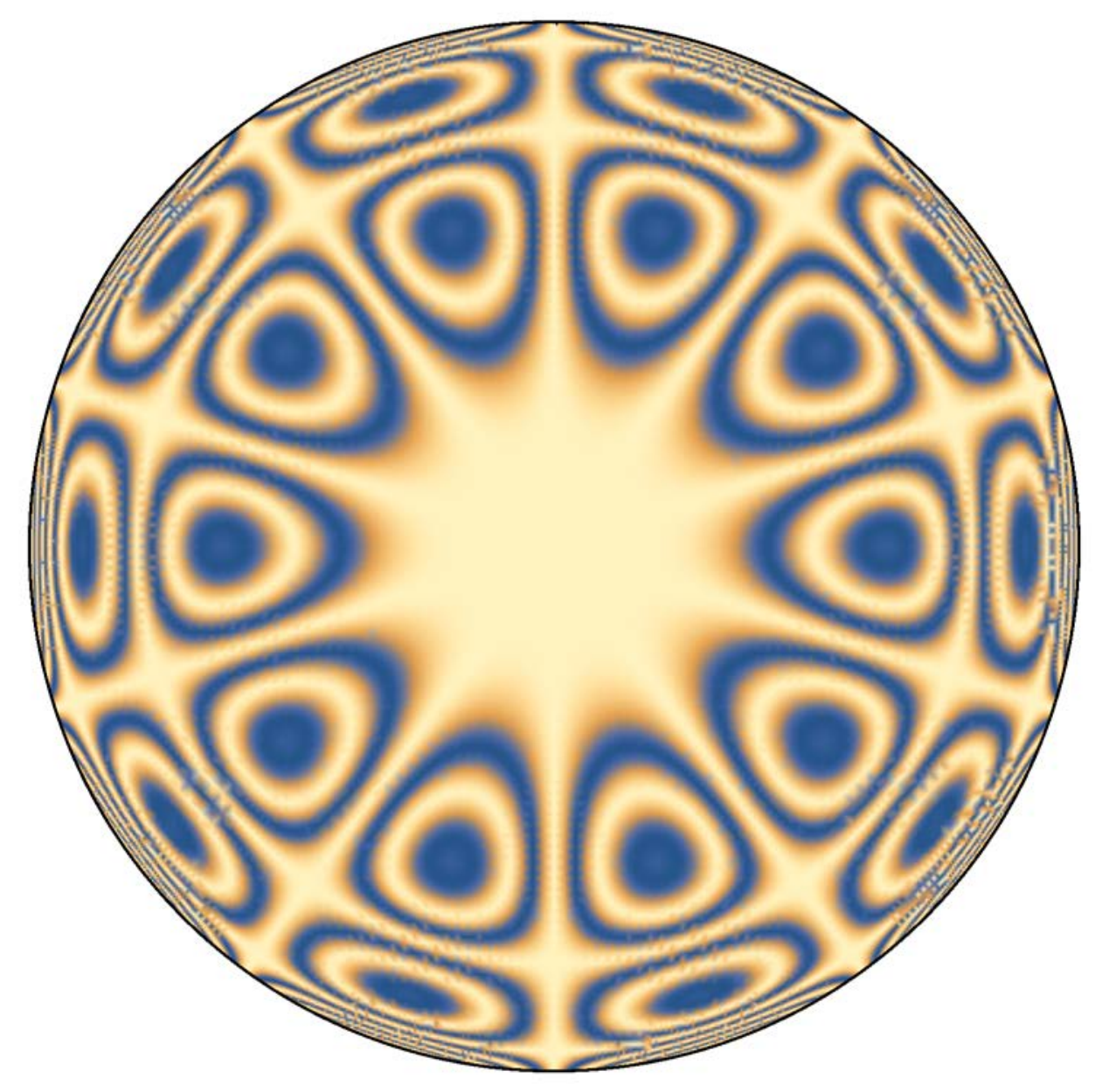}}
\caption{\footnotesize { (a) $W$-Zernike function; (b) Its transformed under $\mathcal O_{\overline\alpha,\overline\beta}$ with $\overline\alpha=(3,0,0)$ and $\overline\beta=(1,0,0)$.}
} 
\label{figure2}
\end{figure}

In Fig.1 we display some 
${\mathcal Z}$-Zernike functions  and their $W$-Zernike counterparts. From the definition \eqref{5} we see that the $W_{u,v}(r,\phi)$ and its counterpart ${\zf}^{u-v}_{u+v}(r,\phi)$
differs in the  factor $\sqrt{{(u+v+1)}/{\pi}}$, whose influence is displayed in Fig.\ref{figure1}. For higher values of $u$ and $v$ the differences are more marked as denoted Figs.1c and 1c$_1$.

In Fig.\ref{figure2} we display $W_{4,1}(r,\phi)$, its transformed under the action of the operator  
$\mathcal O_{\overline\alpha,\overline\beta}$ \eqref{66} where $\overline\alpha=(3,0,0)$ and $\overline\beta=(1,0,0)$, i.e. 
\[
\mathcal O_{\overline\alpha,\overline\beta}\, W_{4,1}(r,\phi)=  {\mathcal A}_+^{3}\,{\mathcal A}_3^{0}\,{\mathcal A}_-^{0}\,{\mathcal B}_+^{1}\,{\mathcal B}_3^{0}\,{\mathcal B}_-^{0}\, 
W_{4,1}(r,\phi)=420\, W_{7,2}(r,\phi)
\]
where the coefficient is computed according  formula \eqref{670}.
 \end{widetext}



\end{document}